\documentclass[12pt]{article}
\usepackage{a4wide}
\usepackage[dvips]{graphicx,psfrag}
\usepackage{amsfonts}
\usepackage{amstext}
\usepackage{amsmath}
\usepackage{amssymb}
\usepackage{amsthm}
\usepackage{cite}
\usepackage[mathscr]{eucal}
\allowdisplaybreaks
\newtheorem{theorem}{Theorem}
\newtheorem{proposition}[theorem]{Proposition}
\newtheorem{lemma}[theorem]{Lemma}
\newtheorem{definition}[theorem]{Definition}
\newtheorem{corollary}[theorem]{Corollary}
\makeatletter
 
 \@addtoreset{equation}{section}
\makeatother
\def\a{\alpha}

\def\Ai{{\rm Ai}}
\def\b{\beta}

\def\C{\mathbb{C}}

\def\e{\epsilon}
\def\E{\mathbb{E}}
\def\F{\mathcal{F}}

\def\G{\mathcal{G}}
\def\i{\infty}

\def\K{\mathcal{K}}
\def\l{\lambda}

\def\1{\bf{1}}
\def\R{\mathbb{R}}
\def\sgn{{\rm{sgn}}}

\def\Z{\mathbb{Z}}
\def\Zm{\mathcal{Z}}

\begin{document}
\title{
Determinantal structures in 
the O'Connell-Yor directed random polymer model
}

\author{Takashi Imamura
\footnote {Department of mathematics and informatics, 
Chiba University,~E-mail:imamura@math.s.chiba-u.ac.jp}
, Tomohiro Sasamoto
\footnote {Department of Physics, Tokyo Institute of Technology,~E-mail: 
sasamoto@phys.titech.ac.jp}}

\maketitle

\begin{abstract}
We study the semi-discrete directed random polymer model 
introduced by O'Connell and Yor. We obtain a representation 
for the moment generating function of the polymer partition function 
in terms of a determinantal measure.
This measure is an extension of the probability measure of the  
eigenvalues for the Gaussian Unitary Ensemble (GUE) in random
matrix theory.
To establish the relation, we introduce another determinantal 
measure on larger degrees of freedom and consider its  few properties, 
from which the representation above follows immediately. 
\end{abstract}

\section{Introduction}
In this paper we consider a directed random polymer model 
in random media in two  (one discrete and one continuous) 
dimension introduced by O'Connell and Yor~\cite{OY2001}.
For $N$ independent one-dimensional standard Brownian motions $B_j(t),~j=1,\cdots,N$ and the 
parameter $\beta (>0)$ representing the inverse temperature, 
the polymer partition function is defined by
\begin{align}
Z_N(t)=\int_{0<s_1<\cdots<s_{N-1}<t}e^{\beta\left(B_1(s_1)+B_2(s_1,s_2)+\cdots+B_N(s_{N-1},t)\right)}ds_1\cdots ds_{N-1}.
\label{ppf}
\end{align}
Here $B_j(s,t)=B_j(t)-B_j(s),~j=2,\cdots,N$ for $s<t$ and
$-B_1(s_1)-B_2(s_1,s_2)-\cdots-B_N(s_{N-1},t)$
represents the energy of the polymer.
In the last fifteen years much progress has been made on this O'Connell-Yor polymer
model, by which we can access some explicit 
information about $Z_N(t)$ and the polymer free energy
$F_N(t)=-\log (Z_N(t))/\beta$~\cite{B2001,BC2014,BCF2012,GTW2001,Ja2014p, Ka2011,Ka2012a,Ka2012b,MO2007,O2012,SV2010}. 
The first breakthrough was made in 
the zero temperature $(\beta\rightarrow\infty)$ case. In this limit, 
$-F_N(t)$ becomes 
\begin{align}
f_N(t):=-\lim_{\beta\rightarrow\infty}F_N(t)
=\max_{0<s_1<\cdots<s_{N-1}<t}
\left(B_1(s_1)+B_2(s_1,s_2)+\cdots+B_N(s_{N-1},t)\right)
\label{ftdef}
\end{align}
where $-f_N(t)$ is the ground state energy. 
For $f_N(t)$, the following relation  
was established~\cite{B2001,GTW2001}:
\begin{align}
&\text{Prob}\left(f_N(t)\le s\right)=\int_{(-\infty,s]^N}\prod_{j=1}^Ndx_j 
\cdot P_{\text{GUE}}(x_1,\cdots,x_N;t),
\label{t0result}
\\
&P_{\text{GUE}}(x_1,\cdots,x_N;t)=
\prod_{j=1}^N\frac{e^{-x_j^2/2t}}{j!t^{j-1}\sqrt{2\pi t}}\cdot
\prod_{1\le j<k\le N}(x_k-x_j)^2,
\label{t0result0}
\end{align}
where $P_{\text{GUE}}(x_1,\cdots,x_N;t)$ is the probability density function of the eigenvalues 
in the Gaussian Unitary Ensemble (GUE) in random matrix 
theory~\cite{AGZ2010,Fo2010,
M2004}. This type of connection of the ground state energy of a
directed polymer in random media with 
random matrix theory
was first obtained for a directed random polymer model on a discrete 
space $\Z_+^2$~\cite{Jo2000} by using the Robinson-Schensted-Knuth(RSK) 
correspondence. 
Eq.~\eqref{t0result} can be regarded as its continuous analogue.
Note that~\eqref{t0result0} is written in the form of  a product of the Vandermonde 
determinant $\prod_{1\le j<k\le N}(x_k-x_j)$. This feature implies that the $m$-point 
correlation function is described by an $m\times m$ determinant, i.e. the eigenvalues 
of the GUE are a typical example of the determinantal point processes~\cite{So2000}.
In addition based on this fact and explicit expression of the correlation kernel, 
we can study the asymptotic behavior of $f_N(t)$ 
in the limit $N\rightarrow\infty$. In~\cite{B2001,GTW2001}, it has been shown that
under a proper scaling, 
the limiting distribution of $f_N(t)$ becomes 
the GUE Tracy-Widom distribution~\cite{TW1994}.

In this paper, we provide a representation for a moment generating function of the polymer partition function~\eqref{ppf}
which holds for arbitrary $\beta(>0)$:
\begin{gather}
\E\left[ \exp\left( -\frac{e^{-\beta u} Z_N(t)}{\beta^{2(N-1)}}  \right) \right]
=\int_{\R^N}\prod_{j=1}^Ndx_j \, f_F(x_j-u)
\cdot W(x_1,\cdots,x_N;t),
\label{result}
\\
W(x_1,\cdots,x_N;t)
=\prod_{j=1}^N\frac{1}{j!}\cdot \prod_{1\le j<k\le N}
(x_k-x_j)\cdot
\det\left(\psi_{k-1}(x_j;t)\right)_{j,k=1}^N,
\label{w1}
\end{gather}
where $f_F(x)=1/(e^{\beta x}+1)$ is the Fermi distribution function and
\begin{align}
\psi_k(x;t)=\frac{1}{2\pi }\int_{-\infty}^{\infty}dw\, e^{-iwx-w^2t/2}
\frac{(iw)^k}{\Gamma\left(1+iw/\beta\right)^N}.
\label{psi1}
\end{align}
For more details see Definition~\ref{defdgue} and Theorem~\ref{thmmain} below.
This is a simple generalization of~\eqref{t0result} to the case of finite temperature.
We easily find that it recovers~\eqref{t0result} in the zero-temperature limit ($\beta\rightarrow\infty$).
Note that the function $W(x_1,\cdots,x_N;t)$ is also written as a product of two determinants
and thus retains the determinantal structure in~\eqref{t0result0}. 

In most cases, to find a finite temperature generalization of results for zero-temperature case is highly nontrivial and in fact often impossible. 
But for the O'Connell-Yor polymer model and a few related models, rich mathematical structures 
have been discovered for finite temperature and the studies on this topic entered a new stage~\cite{ACQ2010,BC2014,COSZ2014,Ha2013,O2012,
OSZ2014, SS2010b,SS2010a,SS2010c,SS2010d}.
In~\cite{O2012}, O'Connell found a connection to the quantum Toda lattice, and 
based on the developments in its studies and the geometric RSK correspondence, 
it was revealed that 
the law of the free energy $F_N(t)$ is expressed as
\begin{align}
\text{Prob}\left(-F_N(t)+\frac{N-1}{\beta}\log\beta^2\le s\right)=
\int_{(-\infty, s]}dx_1\int_{\R^{N-1}}\prod_{j=2}^{N}
dx_{j}\cdot m(x_1,\cdots,x_N;t).
\label{whit0}
\end{align}
Here the probability measure $m(x_1,\cdots,x_N;t)\prod_{j=1}^Ndx_j$, which is called 
the {\it Whittaker measure}, is defined by the density function $m(x_1,\cdots,x_N;t)$  
in terms of the Whittaker function $\Psi_{\lambda}(x_1,\cdots,x_N)$ (for the definition, see~\cite{O2012}) 
and the Sklyanin measure $s_N(\lambda)d\lambda$
(see~\eqref{zsk} below) 
as follows,
\begin{align}
m(x_1,\cdots,x_N;t)
=\Psi_0(\beta x_1,\cdots,\beta x_N)\int_{(i\R)^N}d\lambda\,
\Psi_{-\lambda/\beta}(\beta x_1,\cdots,\beta x_N)e^{\sum_{j=1}^N\lambda_j^2t/2}s_N(\lambda/\beta),
\label{whitm0}
\end{align}
where $\lambda$ represents $(\lambda_1,\cdots,\lambda_N)$.
In contrast to~\eqref{t0result0}, the density function~\eqref{whitm0} is not known to  
be expressed as a product of determinants and 
the process associated with~\eqref{whitm0} does not seem to be determinantal.
Nevertheless some determinantal formulas for the O'Connell-Yor polymer have been found: 
First in~\cite{O2012}, O'Connell showed a determinantal 
representation for the moment generating function (LHS of~\eqref{result}) 
in terms of the Sklyanin measure. (See~\eqref{ocrep} below.)
Next in~\cite{BC2014}, Borodin and Corwin obtained 
a Fredholm determinant representation for the same moment generating 
function (see~\eqref{bcrep} below). A direct proof of the equivalence between the two determinantal
expressions  was given in~\cite{BCR2013}. In~\cite{BC2014}, by considering its continuous limit, 
the authors also obtained an explicit representation of the free energy distribution 
for the directed random polymer in two continuous dimension described 
by stochastic heat equation (SHE)~\cite{BC2014,BCF2012}. 
The distribution in this limit, which describes the universal crossover between 
the Kardar-Parisi-Zhang (KPZ) and the Edwards-Wilkinson
universality class, was first obtained in~\cite{ACQ2010,SS2010b,SS2010a,SS2010c,SS2010d} 
and can be interpreted also as the height distribution for the KPZ
equation~\cite{KPZ1986}.  Furthermore in~\cite{BC2014}, they consider not only the O'Connell-Yor model but 
a class of stochastic processes having the similar Fredholm determinant expressions,  
the Macdonald processes, the probability measures on a sequence of partitions 
which are written in terms of the Macdonald symmetric functions and include the Whittaker measure defined by~\eqref{whit0} 
as a limiting case.

The purpose of this paper is to investigate further the mechanism of appearance of such determinantal
structures and~\eqref{result} is the central formula in our study.  
Although $W(x_1,\cdots,x_N;t)\prod_{j=1}^Ndx_j$ defined by~\eqref{w1}
is not a probability measure but a signed 
measure except when $\beta\rightarrow\infty$,
a remarkable feature of this measure is that  it is {\it determinantal} 
for arbitrary $\beta$ contrary to the Whittaker measure~\eqref{whitm0}.
This determinantal structure allows us to use the conventional techniques
developed in the random matrix theory and thus from the relation 
we readily get a Fredholm determinant representation with a kernel
using biorthogonal functions which is regarded as a generalization of 
the kernel with the Hermite polynomials for the GUE. 
In~\eqref{w1}, the parameter $\beta$, which originally represents the inverse temperature in 
the polymer model appears in the Fermi distribution function $f_F(x-u)$ with the chemical potential $u$ as well as 
$\psi_k(x;t)$~\eqref{psi1} in RHS. This fact with the determinantal structure suggests that the RHS 
might have something to do with the free Fermions at a finite temperature. Related to this,
a curious relation of the height of the KPZ equation with Fermions 
has been discussed in~\cite{DLDMS2015}. 

For establishing the relation,
we introduce a measure on a larger space $\R^{N(N+1)/2}$.
By integrating the measure in two different ways, we get its two marginal weights. 
In one formula appears a determinant which solves the $N$ dimensional diffusion 
equation with some condition
(see \eqref{lemma3},~\eqref{Rdiffusion}, and~\eqref{Rbc})
and the other one with a symmetrization is exactly the RHS of~\eqref{result}. 
The relation~\eqref{result} 
follows immediately from the equivalence of these two expressions.
Our approach is similar to the one by Warren~\cite{Wa2007}
for getting the relation~\eqref{t0result}. 
Actually in the zero-temperature limit
$\beta\rightarrow\infty$, we see that the integration of the measure 
is written in terms of  the probability measure introduced in~\cite{Wa2007}, which describes the positions 
of the reflected Brownian particles on the Gelfand-Tsetlin cone.
Note that the Macdonald processes (especially the Whittaker 
process in our case)~\cite{BC2014} are also another 
generalizations of~\cite{Wa2007}. 
Although the Whittaker process has rich integrable properties, 
they do not inherit the determinantal structure of~\cite{Wa2007}. 
On the other hand, our measure is described without using 
the Whittaker functions and keeps the determinantal structure.
Furthermore combining~\eqref{result} with the fact that the quantity can be rewritten as the 
Fredholm determinant found in \cite{BC2014} (Corollary~\ref{corfred} and Proposition~\ref{propfred2} below), 
our approach can be considered as another proof of the equivalence between~\eqref{bcrep} 
and~\eqref{ocrep} in~\cite{BCR2013}. One feature of our proof is to bring to light the larger determinantal structure
behind the two relations. 

This paper is organized as follows. In the next section, after stating
the definition of a determinantal measure, 
we give our main result, Theorem~\ref{thmmain} and its proof. 
The proof consists of two major steps:
we first  introduce in Lemma~\ref{lemOC} a determinantal representation for the moment generating function which is a deformed version of the representation~\eqref{ocrep} in~\cite{O2012}. 
Next we introduce another determinantal measure on larger space $\R^{N(N+1)/2}$ 
and then we find two relations about its integrations which play a key role in deriving our main result.
In Sec. 3 we show that
this approach can be considered as an extension of the one in~\cite{Wa2007}. 
In Sec.4., we consider the Fredholm determinant formula with biorthogonal kernel 
obtained by applying conventional 
random matrix techniques to our main result.
The scaling limit to the KPZ equation is discussed in Sec.5. We check that 
our kernel goes to the one obtained in the studies of the KPZ equation.
A concluding remark is given in the last section.

\section{Main result}\label{MR}
In this section, we introduce a measure $W(x_1,\cdots,x_N;t)\prod_{j=1}^Ndx_j$~\eqref{w1}, 
state our main result  and give its proof.

\subsection{Definition and result}
\begin{definition}\label{defdgue}
Let 
$\psi_k(x;t),~k=1,2,\cdots$ be
\begin{align}
\psi_k(x;t)=\frac{1}{2\pi }\int_{-\infty}^{\infty}dw\, e^{-iwx-w^2t/2}
\frac{(iw)^k}{\Gamma\left(1+iw/\beta\right)^N}.
\label{psi}
\end{align}

For $(x_1,\cdots,x_N)\in\R^N$, a function $W(x_1,\cdots,x_N;t)$
is defined by
\begin{align}
W(x_1,\cdots,x_N;t)=\prod_{j=1}^N\frac{1}{j!}\cdot \prod_{1\le l<m\le N}(x_m-x_l)\cdot
\det\left(\psi_{j-1}(x_k;t)\right)_{j,k=1}^N.
\label{w}
\end{align}
\end{definition}
\noindent
{\bf Remark.}
We find that 
$W(x_1,\cdots,x_N;t)$
is a real function on $\R^N,$
since by definition $\psi_k(x;t)$ is real for any 
$k=0,1,2,\cdots, N-1, \beta>0$ and $t>0$.  
But in general, the positivity of this measure is not guaranteed.
For example $\psi_0(x;t)$ shows a damped oscillation and  
can take a negative value for some $x$. Thus at least
for the case $N=1$, $W(x,t)=\psi_0(x;t)$ can be negative.

\vspace{2mm}

We discuss the zero-temperature limit  $\beta\rightarrow \i$
of $W(x_1,\cdots,x_N;t)$.
Noting 
$\Gamma(1)=1$, 
we see  
\begin{align}
&\lim_{\beta\rightarrow\infty}\psi_k(x;t)=\frac{1}{2\pi}
\int_{-\infty}^{\infty}dw\, e^{-iwx-w^2t/2}(iw)^k
=\frac{e^{-x^2/2t}}{\sqrt{2\pi t}}
\left(\frac{1}{2t}\right)^{\frac{k}{2}}
H_k\left(\frac{x}{\sqrt{2t}}\right),
\label{psibii}
\end{align}
where we used the integral representations of the $n$th order Hermite polynomial $H_n(x)$
(see e.g. Section 6.1 in~\cite{AAR1999}),
\begin{align}
H_n(x)=\frac{(-2i)^n}{\sqrt{\pi}}
\int_{-\infty}^{\infty}du\, u^ne^{-(u-ix)^2}
\label{hermite}
\end{align}
Note that $(t/2)^{k/2}H_k(x/\sqrt{2t})$ is a monic polynomial (i.e.
the coefficient of the highest degree is 1) and
\begin{align}
\lim_{\beta\rightarrow\infty}\det\left(\psi_{k-1}(x_j;t)\right)_{j,k=1}^N
=\prod_{j=1}^N\frac{e^{-x_j^2/2t}}{t^{j-1}\sqrt{2\pi t}}\cdot\prod_{1\le j<k\le N}(x_k-x_j).
\label{dpsilim}
\end{align}
Thus we find
\begin{align}
\lim_{\beta\rightarrow\infty}W(x_1,\cdots,x_N;t)=
P_{\text{GUE}}(x_1,\cdots,x_N;t),
\label{lGUE}
\end{align}
where $P_{\text{GUE}}(x_1,\cdots,x_N;t)$ is defined by~\eqref{t0result0}. 
The function $W(x_1,\cdots,x_N;t)$ can 
be regarded as a
deformation of ~\eqref{t0result0} which keeps its determinantal structure.

In this paper, we provide a determinantal representation for
the moment generating function 
of the polymer partition function~\eqref{ppf} in terms of the
function~\eqref{w}.
\begin{theorem}\label{thmmain}
\begin{align}
\E\left(e^{-\frac{e^{-\beta u} Z_N(t)}{\beta^{2(N-1)}}}\right)
=\int_{\R^N}\prod_{j=1}^Ndx_j
\,
f_F(x_j-u)
\cdot W(x_1,\cdots,x_N;t)
\label{main}
\end{align}
where 
$f_F(x)=1/(e^{\beta x}+1)$ is the Fermi distribution function.
\end{theorem}

By~\eqref{ftdef},~\eqref{lGUE} and the simple facts 
\begin{align}
\lim_{\beta\rightarrow\infty}e^{-e^{\beta x}}=\lim_{\beta\rightarrow x}f_F(x)=\Theta (-x),
\end{align}
we find that the zero temperature limit of~\eqref{main} becomes~\eqref{t0result}.

Because of the determinantal structure of $W(x_1,\cdots,x_N;t)$, we can get the Fredholm determinant
representation for the moment generating function by using the techniques in random matrix theory.
Recently another Fredholm determinant representation has been given based on properties of Macdonald 
difference operators~\cite{BC2014}.  The equivalence between them 
will be shown in Sec.~\ref{fdf}. 

\subsection{Proof}\label{Proof}
Here we provide a proof of Theorem~\ref{thmmain}.
Our starting point is the representation
for the moment generating function given in~\cite{O2012}:
\begin{align}
\E\left(e^{-\frac{e^{-\beta u} Z_N(t)}{\beta^{2(N-1)}}}\right)
=\int_{(i\R-\e)^N}
\prod_{j=1}^N
\frac{d\l_j}{\beta}e^{-u\lambda_j+\lambda_j^2t/2}\Gamma\left(-\frac{\lambda_j}{\beta}\right)^N\cdot s_N\left(\frac{\lambda}{\beta}\right),
\label{ocrep}
\end{align}
where $0<\e<\beta$ and $s_N(\lambda)d\lambda$ is the Sklyanin measure defined by
\begin{align}
s_N(\lambda)=\frac{1}{(2\pi i)^NN!}
\prod_{i<j}\frac{\sin\pi (\lambda_i-\lambda_j)}{\pi}
\prod_{i>j}\left(\lambda_i-\lambda_j\right). 
\label{zsk}
\end{align}
This relation was obtained by using  
the properties of the Whittaker functions~\cite{Bu1989,St2002}
and the Whittaker measure~\eqref{whitm}.

\begin{lemma}\label{lemOC}
\begin{align}
\E\left(e^{-\frac{e^{-\beta u} Z_N(t)}{\beta^{2(N-1)}}}\right)
=
\int_{\R^N}\prod_{\ell=1}^N dx_\ell f_F(x_\ell-u)\cdot
G(x_1,\cdots,x_N;t)
\label{lemma3}
\end{align}
where $f_F(x)$ is defined below~\eqref{cor4} and
\begin{align}
&G(x_1,\cdots,x_N;t)=\det\left(F_{jk}(x_{N-j+1};t)\right)_{j,k=1}^N,
\label{fG}
\\
&
F_{jk}(x;t)=\int_{i\R-\e}\frac{d\lambda}{2\pi i}
\frac{e^{-\lambda x+\lambda^2 t/2}}{\Gamma\left(\frac{\lambda}{\beta}+1\right)^N}
\left(\frac{\pi}{\beta}\cot\frac{\pi\lambda}{\beta}\right)^{j-1}
\lambda^{k-1}
\label{fjk}
\end{align}
with $0<\e<\beta$.
\end{lemma}

\smallskip
We will discuss an interpretation of~\eqref{fG} 
in the next section.  In this definition, we have arranged $x_i$\rq{}s in
the reversed order so as to relate~\eqref{Wf}, the zero-temperature limit of \eqref{fG}, 
to the stochastic processes defined later in~\eqref{refBM}. 

\smallskip
\noindent
{\bf Proof.}
Noting the relation
\begin{align}
&~\prod_{1\le i<j\le N}{\sin(x_i-x_j)}
=
\prod_{1\le i<j\le N}\sin x_i \sin x_j \left(\cot x_j-\cot x_i\right)\notag\\
&=\prod_{j=1}^N\sin^{N-1}x_j\cdot
\prod_{1\le k<\ell\le N}\left(\cot x_\ell-\cot x_k\right)
=\prod_{j=1}^N\sin^{N-1}x_j\cdot\det\left(\cot^{\ell-1}x_k
\right)_{k,\ell=1}^N,
\end{align}
we rewrite RHS of~\eqref{ocrep} as
\begin{align}
&~~\int_{(i\R-\e)^N}\prod_{j=1}^N
\frac{d\l_j}{\beta}e^{-u\lambda_j+\lambda_j^2t/2}\Gamma\left(-\frac{\lambda_j}{\beta}\right)^N\cdot s_N\left(\frac{\lambda}{\beta}\right)\notag\\
&=
\frac{1}{N!}\int_{(i\R-\e)^N}
\prod_{j=1}^N\frac{d\lambda_j}{2\pi i\beta}e^{-u\lambda_j+\lambda_j^2t/2}
\Gamma\left(-\frac{\lambda_j}{\beta}
\right)^N\left(\frac{\sin\frac{\pi}{\beta}\lambda_j}{\pi}\right)^{N-1}\notag\\
&\hspace{1.8cm}\times
\det\left(\left(\frac{\pi}{\beta}\cot\frac{\pi}{\beta}\lambda_j\right)^{k-1}\right)_{j,k=1}^N
\det\left({\lambda_j}^{k-1}\right)_{j,k=1}^N
\notag\\
&=\det\left(\int_{i\R-\e}
\frac{d\lambda}{2\pi i\beta}e^{-u\lambda+\lambda^2t/2}
\Gamma\left(-\frac{\lambda}{\beta}
\right)^N\left(\frac{\sin\frac{\pi}{\beta}\lambda}{\pi}\right)^{N-1}
\left(\frac{\pi}{\beta}\cot\frac{\pi}{\beta}\lambda\right)^{j-1}
\lambda^{k-1}
\right)_{j,k=1}^N
\label{re1}
\end{align}
where in the last equality, we used the Andr{\'e}ief identity (also known as
the Cauchy-Binet identity)~\cite{And1883}:
For the functions $g_j(x),~h_j(x)$, $j=1,2,\dots,N,$ such that all integrations below are 
well-defined, we have
\begin{align}
\frac{1}{N!}\int_{\R^N}\prod_{j=1}^Ndx_j\cdot\det\left(g_k(x_j)\right)_{j,k=1}^N\det\left(h_k(x_j)\right)_{j,k=1}^N=\det\left(\int_{\R}dx g_j(x)h_k(x)\right)_{j,k=1}^N.
\label{heine}
\end{align}

We notice that the factor $e^{-u\lambda}\Gamma(-\lambda/\beta)^
N({\sin(\pi\lambda/\beta)}/{\pi})^{N-1}$in~\eqref{re1}
can be written as
\begin{align}
e^{-u\lambda}\Gamma\left(-\frac{\lambda}{\beta}\right)^N
\left(\frac{\sin\frac{\pi}{\beta}\lambda}{\pi}\right)^{N-1}
&=\frac{(-1)^{N-1}}{\Gamma\left(1+\frac{\lambda}{\beta}\right)^N}
\frac{\pi e^{-u\lambda}}{-\sin\frac{\pi}{\beta}\lambda}\notag\\
&=
\frac{(-1)^{N-1}}{\Gamma\left(1+\frac{\lambda}{\beta}\right)^N}
\int_{-\infty}^{\infty}\beta \frac{e^{-x\lambda}}{e^{\beta(x-u)}+1}dx
\label{re2}
\end{align}
where we used the reflection formula for the Gamma function and 
the relation~\eqref{sinrelation}.
From~\eqref{re1} and~\eqref{re2}, 
we arrive at the desired expression~\eqref{lemma3}.
\qed

From (\ref{lemma3}), we see that for the derivation of our main result (\ref{main}), 
it is sufficient to prove the relation
\begin{align}
\int_{\R^N} 
\prod_{\ell=1}^N dx_\ell f_F(x_\ell-u)\cdot
G(x_1,\cdots,x_N;t)
=\int_{\R^N}\prod_{j=1}^Ndx_j\, f_F(x_j-u)
\cdot W(x_1,\cdots,x_N;t).
\label{trelation}
\end{align}
where $f_F(x)$ is defined below~\eqref{main} and $W(x_1,\cdots, x_N;t)$
is given in Definition~\ref{defdgue}. 
Note that this is a relation for the integrated values on $\R^N$.
To establish this 
we introduce a measure on the larger space $\R^{N(N+1)/2}$. 

\begin{definition}\label{deftr}
Let $\underline{x}_k$ be an array $(x^{(1)},\cdots, x^{(k)})$ where $x^{(j)}=(x^{(j)}_1,\cdots,x^{(j)}_j)\in\R^{j}$
and $d\underline{x}_k=\prod_{j=1}^k\prod_{i=1}^jdx^{(j)}_i$.
We define a measure
$R_u(\underline{x}_N;t)d\underline{x}_N$ 
by
\begin{align}
R_u(\underline{x}_N;t)=\prod_{1\le i\le j\le N} 
f_i(x^{(j)}_i-x^{(j-1)}_{i-1})
\cdot\det\left(F_{1i}(x^{(N)}_j;t)\right)_{i,j=1}^N.
\label{taweight}
\end{align}
Here $x^{(j-1)}_0=u$, $F_{1j}(x;t)$ is given by $F_{ij}(x;t)$~\eqref{fjk} with $i=1$ and 
$f_i(x), i=1,2,\cdots$ is  defined by using the Fermi and Bose distribution functions,
$f_F(x):=1/(e^{\beta x}+1)$ and $f_B(x):=1/(e^{\beta x}-1)$ respectively as follows.
\begin{align}
f_i(x)=
\begin{cases}
f_F(x), & i=1,\\
f_B(x), & i\ge 2.
\end{cases}
\label{ffB}
\end{align}
\end{definition}

\vspace{2mm}
\noindent
{\bf Remark.}
The reason why both the Bose and Fermi distributions appear in our approach 
is not clear. The interrelations between them (see~\eqref{convdef}-\eqref{r2} below) will play an
important role in the following discussions.    

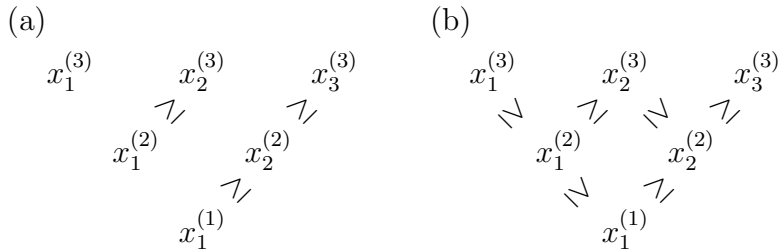
\begin{figure}[h]
\begin{picture}(350,90)
\put(90, 80){(a)}
\put(105,60){$x^{(3)}_1$}
\put(155,60){$x^{(3)}_2$}
\put(205,60){$x^{(3)}_3$}
\put(145,45){\rotatebox{45}{$\ge$}}
\put(195,45){\rotatebox{45}{$\ge$}}
\put(130,30){$x^{(2)}_1$}
\put(180,30){$x^{(2)}_2$}
\put(170,15){\rotatebox{45}{$\ge$}}
\put(155,0){{$x^{(1)}_1$}}
\put(250,80){(b)}
\put(265,60){$x^{(3)}_1$}
\put(315,60){$x^{(3)}_2$}
\put(365,60){$x^{(3)}_3$}
\put(275,50){\rotatebox{-45}{$\ge$}}
\put(305,45){\rotatebox{45}{$\ge$}}
\put(330,50){\rotatebox{-45}{$\ge$}}
\put(355,45){\rotatebox{45}{$\ge$}}
\put(290,30){$x^{(2)}_1$}
\put(340,30){$x^{(2)}_2$}
\put(300,20){\rotatebox{-45}{$\ge$}}
\put(330,15){\rotatebox{45}{$\ge$}}
\put(315,0){{$x^{(1)}_1$}}
\end{picture}
\caption{\label{vtgt} Triangular arrays ($k=3$) (a) an element of $V_k$~\eqref{vn}
(b) The Gelfand-Tsetrlin pattern (an element of~\eqref{gtn})
}
\end{figure}

As in Fig~\ref{vtgt}. we usually represent the array $\underline{x}_N$ graphically in 
the triangular shape. Although no ordering is imposed on  $\underline{x}_N$,
in the zero-temperature limit, $R_u(\underline{x}_N;t)$
has the support on the ordered arrays as in~Fig.~\ref{vtgt} (a) (see~\eqref{vn}).
Fig.~\ref{vtgt} (b) represents the other ordered array called the Gelfand-Tsetlin pattern
(see~\eqref{gtn}).

As discussed later we will find that  the moment generating function of the 
O'Connell-Yor polymer model is expressed as the integration of this measure 
$R_u(\underline{x}_N;t)$ over $\R^{N(N+1)/2}$. We have other choices for 
the definition of $R_u(\underline{x}_N;t)$ which give the same integration value.
One example 
is 
\begin{align}
\bar{R}_u(\underline{x}_N;t)=
\prod_{\ell=1}^N\frac{1}{\ell!}
\det\left(f_i(x^{(\ell)}_j-x^{(\ell-1)}_{i-1})\right)_{i,j=1}^\ell
\cdot\det\left(F_{1i}(x^{(N)}_j;t)\right)_{i,j=1}^N.
\label{taweightb}
\end{align}
This comes form the following consideration. Let $f_{\text{sym}}(\underline{x}_N)$ be
a function which is symmetric under permutations of 
$x^{(j)}_1,\cdots,x^{(j)}_j$ for each $j\in \{1,2,\cdots,N\}$. Then we see that 
$R_u(\underline{x}_N;t)$~\eqref{taweight} and $\bar{R}_u(\underline{x}_N;t)$ have the
same integration value:
\begin{align}
\int_{\R^{N(N+1)/2}}d\underline{x}_N
f_{\rm{sym}}(\underline{x}_N)R_u(\underline{x}_N;t)
=
\int_{\R^{N(N+1)/2}} d\underline{x}_N
f_{\rm {sym}}(\underline{x}_N)\bar{R}_{u}(\underline{x}_N;t).
\label{rrbar}
\end{align}
It can be shown as follows. 
From the symmetry of $f_{\text{sym}}(\underline{x}_N)$, LHS of the equation above becomes
\begin{align}
\int_{\R^{N(N+1)/2}}d\underline{x}_N
f_{\rm{sym}}(\underline{x}_N)R_u(\underline{x}_N;t)=
\int_{\R^{N(N+1)/2}}d\underline{x}_N
f_{\rm{sym}}(\underline{x}_N)\tilde{R}_u(\underline{x}_N;t).
\label{Rusym}
\end{align}
Here $\tilde{R}_u(\underline{x}_N;t)$ is defined by
\begin{align}
\tilde{R}_u(\underline{x}_N;t)=\prod_{\ell=1}^N\frac{1}{\ell!}\sum_{\sigma^{(j)}\in S_j,~j=1,\cdots,N}
R_u\left(\underline{x}_N^\sigma;t\right),
\label{tru}
\end{align}
where $S_j$ is the permutation of $1,2,\cdots,j$ and $\underline{x}_N^\sigma$ denotes
$(x^{\sigma^{(1)}},\cdots, x^{\sigma^{(N)}})$ with $x^{\sigma^{(j)}}=(x^{(j)}_{\sigma^{(j)}(1)},\cdots,
x^{(j)}_{\sigma^{(j)}(j)})$. We easily find the equivalence 
$\tilde{R}_u(\underline{x}_N;t)=\bar{R}_u(\underline{x}_N;t)$.  Note that
\begin{align}
&~R_{u}(\underline{x}_N^\sigma;t)
=
\prod_{1\le i\le j\le N} f_i\left(x^{(j)}_{\sigma^{(j)}(i)}-x^{(j-1)}_{\sigma^{(j-1)}(i-1)}\right)
\cdot\det\left(F_{1i}(x^{(N)}_{\sigma^{(N)}(j)};t)\right)_{i,j=1}^N\notag\\
&
=\sgn\sigma^{(N)}\prod_{j=1}^N\prod_{i=1}^{j} f_i\left(x^{(j)}_{\sigma^{(j)}(i)}-x^{(j-1)}_{\sigma^{(j-1)}(i-1)}\right)
\cdot\det\left(F_{1i}(x^{(N)}_{j};t)\right)_{i,j=1}^N\notag\\
&
=\prod_{j=1}^N\sgn
\tau^{(j)}
\prod_{i=1}^{j} f_i\left(x^{(j)}_{\tau^{(j)}(i)}-x^{(j-1)}_{i-1}\right)
\cdot\det\left(F_{1i}(x^{(N)}_{j};t)\right)_{i,j=1}^N.
\label{xrho}
\end{align}
Here in the last equality, $\tau^{(j)}\in S_j,~j=1,2,\cdots,N$ is defined
by using $\sigma^{(j-1)}$ and $\sigma^{(j)}$ as
$\sigma^{(j-1)}\tau^{(j)}(k)=\sigma^{(j)}(k),~k=1,\cdots,j$, where we regard
$\sigma^{(j-1)}$ as an element of $S_j$ with $\sigma^{(j-1)}(j)=j$.
Further in the last equality we used $\sigma^{(N)}=\prod_{j=1}^N\tau^{(j)}$. Substituting~\eqref{xrho} into~\eqref{tru} and using the definition
of the determinant, we have $\tilde{R}_u(\underline{x}_N;t)=\bar{R}_u(\underline{x}_N;t)$.

The function $\bar{R}_u(\underline{x}_N;t)$~\eqref{taweightb} has a similar determinantal structure 
to the Schur process~\cite{OkRe2003}.
The Schur process is a probability measure on the sequence of partitions $\{\lambda^{(j)}\}_{j=1,\cdots,N},$
where  $\lambda^{(j)}:=\{(\lambda^{(j)}_1,\cdots,\lambda^{(j)}_j)|\lambda^{(j)}_i\in \Z,~\lambda^{(j)}_1
\ge\cdots\ge\lambda^{(j)}_j\ge 0\}$, described as products of the skew Schur functions $s_{\lambda/\mu}(x_1,\cdots,x_n)$.
For the ascending case (see Definition 2.7 in~\cite{BC2014}), the probability measure is expressed as
\begin{align}
\prod_{i,j=1}^N\frac{1}{1-a_ib_j}\cdot\prod_{k=1}^{N}s_{\lambda^{(k)}/\lambda^{(k-1)}} (a_k)\cdot s_{\lambda^{(N)}}(b_1,\cdots, b_N),
\label{asp}
\end{align} 
where $a_j,~b_j,~j=1,\cdots,N$ are positive variables.
We note that 
$s_{\lambda^{(k)}/\lambda^{(k-1)}} (a_k)$ is expressed as a $k$th order 
determinant and  $s_{\lambda^{(N)}}(b_1,\cdots, b_N)$
as a $N$th order determinant 
by the Jacobi-Trudi identity~\cite{Ma1995}, 
\begin{align}
s_{\lambda/\mu}(x_1,\cdots,x_n)=\det\left(h_{\lambda_i-\mu_j+j-i}(x_1,\cdots,x_n)\right)_{i,j=1}^{\ell(\lambda)},
\end{align}
where $h_{k}(x_1,\cdots,x_n)$ is a complete symmetric polynomial with degree $k$ and $\ell(\lambda)$
is the length of the partition $\lambda$.
Thus~\eqref{taweightb} and~\eqref{asp} have a common
structure of $N$ products of determinants with increasing size times an $N$th order
determinant.

\vspace{3mm}

In the following we provide the relations about two marginals of $R_u(\underline{x}_N;t)$~\eqref{taweight},
from which~\eqref{trelation} immediately follows. For this purpose, we give 
two formulas for $f_F(x)$ and $f_B(x)$~\eqref{ffB}.
First we define a multiple convolution $g^{*(m)}f(x)~m=0,1,2,\cdots$ for 
a functions $f(x)$ on $\R$ and an integral operator  $g$ with the kernel $g(x-y)$
as
\begin{align}
g^{*(0)}f(x)=f(x),~g^{*(k)}f(x)=\int_{-\infty}^{\infty} dy\, g(x-y)g^{*(k-1)}f(y),~k=1,2,\cdots.
\label{convdef} 
\end{align}
Using this definition, the formulas are written as follows:
\begin{lemma} \label{lem9}
We regard all integrations below 
as the Cauchy principal values. 
For $\beta>0$, $a\in\C$ with $-\beta<\text{Re~}a<0$ and $m=0,1,2,\cdots$, we have
\begin{align}
&f^{*(m)}_Be^{ax}
=\left(\frac{\pi}{\beta} \cot \left(\frac{\pi a}{\beta} \right)\right)^{\hspace{-1mm}m} e^{ax},
\label{r1}\\
&f^{*(m)}_Bf_F(x)
=q_m(x)f_F(x),
\label{r2}
\end{align}
where $q_m(x)$ is an $m$th order polynomial with
the coefficient of the highest degree being $1/m!$.
\end{lemma}

\noindent
A proof of this lemma will be given in Appendix~\ref{B}.
The polynomial $q_m(x)$ in~\eqref{r2}  is defined inductively 
by~\eqref{irec}-\eqref{jbjf}. But in our later discussion we will not use 
its explicit form.

From~\eqref{fjk} and~\eqref{r1},
we readily obtain for $m=0,1,2\cdots$,
\begin{align}
\tilde{f}_B^{*(m)}F_{jk}\left(x;t\right)=F_{j+m,k}(x;t),
\label{th93}
\end{align}
where we define $\tilde{f}_B(x):=f_B(-x)$.

Using~\eqref{r2} and~\eqref{th93}, we obtain the following relations.
\begin{theorem}\label{mar}
Let the measures $dA_1$ and $dA_2$ be
\begin{align}
dA_1=\prod_{2\le i\le j\le N}d x^{(j)}_i,~dA_2=\prod_{1\le i\le j\le N-1}dx^{(j)}_i.
\label{dAdB}
\end{align}
Then we have
\begin{align}
&\int_{\R^{N(N-1)/2}}dA_1 R_u(\underline{x}_N;t)
=
G(x^{(1)}_1,\cdots,x^{(N)}_1;t)\prod_{j=1}^N f_F(x^{(j)}_1-u),
\label{dA1}
\\
&
\int_{\R^{N(N-1)/2}}dA_2 R_u(\underline{x}_N;t)
=\bar{W}(x^{(N)}_1,\cdots,x^{(N)}_N;t)\prod_{j=1}^N f_F(x^{(N)}_j-u).
\label{dA2}
\end{align}
Here $G(x^{(1)}_1,\cdots,x^{(N)}_1;t)$ is defined by~\eqref{fG} and
\begin{align}
\bar{W}(x^{(N)}_1,\cdots,x^{(N)}_N;t)=\prod_{j=1}^{N-1}
q_j\left(x^{(N)}_{j+1}-u\right)
\cdot 
\det\left(F_{1j}\left(x^{(N)}_k;t\right)\right)_{j,k=1}^N,
\end{align}
where $q_j(x)$ is defined below~\eqref{r2}.
\end{theorem}

\smallskip

We easily see that~\eqref{trelation} can be obtained from these relations~\eqref{dA1}
and~\eqref{dA2}: Integrating the both hand sides of them over the remaining degrees of 
freedom ($(x^{(1)}_1,\cdots, x^{(N)}_1)$ for~\eqref{dA1} and $(x^{(N)}_1,\cdots,x^{(N)}_N)$ 
for~\eqref{dA2}),
we get two different expression about the integrated value of 
$R_u(\underline{x}_N;t)$
\begin{align}
&\int_{\R^{N(N+1)/2}}d\underline{x}_N R_u(\underline{x}_N;t)
=\int_{\R^N}\prod_{j=1}^N dx^{(j)}_1f_F(x^{(j)}_1-u)\cdot
G(x^{(1)}_1,\cdots,x^{(N)}_1;t)
\label{idA1}
\\
&
\int_{\R^{N(N+1)/2}}d\underline{x}_N R_u(\underline{x}_N;t)
=\int_{\R^N}\prod_{j=1}^N dx^{(N)}_j f_F(x^{(N)}_j-u)\cdot\bar{W}(x^{(N)}_1,\cdots,x^{(N)}_N;t).
\label{idA2}
\end{align}
where $d\underline{x}_N=\prod_{1\le i\le j\le N}x^{(j)}_i$.
RHS of the second relation is further rewritten as
\begin{align}
\int_{\R^N}\prod_{j=1}^Ndx^{(j)}_1f_F(x^{(j)}_1-u)\cdot\frac{1}{N!}
\sum_{\sigma^{(N)}\in S_N}\bar{W}(x_{\sigma^{(N)}(1)},\cdots,x_{\sigma^{(N)}(N)};t),
\label{ggsym1}
\end{align}
and the symmetrized  $\bar{W}(x_1,\cdots,x_N;t)$ in this equation
is nothing but $W(x^{(N)}_1,\cdots,x^{(N)}_1;t)$~\eqref{w} since
\begin{align}
\frac{1}{N!}\sum_{\sigma^{(N)\in S_N}}\bar{W}(x_{\sigma^{(N)}(1)},\cdots,x_{\sigma^{(N)}(N)};t)
&=\frac{1}{N!}\cdot\det\left(q_{j-1}(x^{(N)}_k)\right)_{j,k=1}^N\det\left(F_{1j}(x^{(N)}_k;t)\right)_{j,k=1}^N\notag\\
&=W(x^{(N)}_1,\cdots,x^{(N)}_N;t).
\label{align} 
\end{align}
Here in the second equality we used the fact that $q_{j}(x)$ is a $j$th order polynomial
with the coefficient of the highest degree being $1/j!$ and $F_{1j}(x;t)=\psi_{j}(x;t)$.


\smallskip
\noindent
{\bf Proof of Theorem~\ref{mar}.}
First we derive \eqref{dA1}.
By the definition of~\eqref{taweight}, LHS of~\eqref{dA1} becomes
\begin{align}
\prod_{j=1}^N f_F(x^{(j)}_1-u)
\cdot\det\left(\tilde{f}_B^{*(k-1)}F_{1j}\left(x^{(N-k+1)}_1;t\right)\right)_{j,k=1}^N.
\label{th91}
\end{align}
Here $\tilde{f}_B(x)$ is defined below~\eqref{th93}. Applying~\eqref{th93} to this equation
we obtain~\eqref{dA1}.

Next we derive \eqref{dA2}. 
We see that the factor $dA_2\prod_{1\le i\le j\le N}f_i(x^{(j)}_i-x^{(j-1)}_{i-1})$
in $R_u(\underline{x_N};t)$~\eqref{taweight}
can be decomposed to
\begin{align}
dA_2\prod_{1\le i\le j\le N} f_{i}\left(x^{(j)}_{i}-x^{(j-1)}_{i-1}\right)
=\prod_{k=1}^{N-1}
\left(\prod_{i=1}^{N-k}dx^{(i+k-1)}_i\cdot\prod_{j=1}^{N-k+1}
f_j\left(x^{(j+k-1)}_j-x^{(j+k-2)}_{j-1}\right)
\right),
\label{bdec2}
\end{align}
and from~\eqref{r2} the integration of the factor for each $k$ is represented as
\begin{align}
&\int_{\R^{N-k}}\prod_{1\le i\le N-k}dx^{(i+k-1)}_i \prod_{j=1}^{N-k+1}f_j(x^{(j+k-1)}_i-x^{(j+k-2)}_{j-1})
=f^{*(N-k)}_Bf_F(x^{(N)}_{N-k+1}-u)\notag\\
&=q_{N-k}(x^{(N)}_{N-k+1}-u)f_F(x^{(N)}_{N-k+1}-u)
\end{align}
where $q_m(x)$ is given in~\eqref{r2}. Eq.~\eqref{dA2} follows immediately from this relation.
\qed


\section{Dynamics of the two marginals}

The purpose of this section is to have a better understanding of the two quantities, 
$W(x_1,\cdots,x_N;t)$~\eqref{w} and $G(x_1,\cdots,x_N;t)$~\eqref{fG}, which
arose as partially integrated quantities of  $R_u(\underline{x}_N;t)$~\eqref{taweight} 
in Theorem~\ref{mar} (for $W$ a symmetrization is also necessary, see (\ref{align})).
We will first consider the evolution equations of these two quantities. 
Next we will see that the zero-temperature limit of the equation for 
$W(x_1,\cdots,x_N;t)$ is nothing but the evolution equation for the Brownian particles 
with reflection interaction while $W(x_1,\cdots,x_N;t)$
satisfies the one for the GUE Dyson's Brownian motion~\cite{Dys1962}
regardless of the value of $\beta$. Furthermore we will find that 
our idea using  $R_u(\underline{x}_N;t)$ in an enlarged space $\R^{N(N+1)/2}$ (Theorem~\ref{mar})  
is similar to the argument in~\cite{Wa2007} although we need a modification
of ~\cite{Wa2007} about the ordering in an enlarged space.

%
%
%
%
%
%
%
%

\subsection{Evolution equations of  $G(x_1,\cdots,x_N;t)$ and $W(x_1,\cdots,x_N;t)$}


Let us first summarize the properties of  $F_{jk}(x;t)$~$j,k\in\{1,2,\cdots\}$~\eqref{fjk}
all of which are easily confirmed by simple observations:
\begin{gather}
F_{1k}(x;t)=\psi_k(x;t),
\label{ffp1}
\\
\frac{\partial}{\partial t}F_{jk}(x;t)=\frac{1}{2}\frac{\partial^2}{\partial x^2}F_{jk}(x;t),
\label{ffp2}
\\
\int_{-\infty}^\infty dx \tilde{f}_B(x-y) F_{jk}(x;t)=F_{j+1k}(y;t), 
\label{ffp3}
\\
-\frac{\beta^2}{\pi^2}\int_{-\infty}^\infty dx_{j+1}\frac{e^{\frac{\beta}{2}(x_{j+1}-x_j)}}
{e^{{\beta}(x_{j+1}-x_j)}-1}F_{j+1k}(x_{j+1};t)=F_{jk}(x_j,k),
\label{ffp4}
\end{gather}
where $\psi_k(x;t)$ in~\eqref{ffp1} and $\tilde{f}_B(x)$ in~\eqref{ffp3} are 
defined by~\eqref{psi} and below~\eqref{th93}. 
Eq.~\eqref{ffp3} is equivalent to
\eqref{th93} while \eqref{ffp4} is obtained from the relation
\begin{align} 
\frac{\beta^2}{\pi^2}\int_{-\infty}^{\infty}dy\,\frac{e^{\frac{\beta}{2}(y-x)}}{e^{\beta (y-x)}-1}e^{-by}
=-\frac{\beta}{\pi}
\tan\left(\frac{\pi}{\beta}b\right)e^{-bx},
\label{ffp5}
\end{align}
for $|\text{Re}~b|<\beta/2$. This relation
is easily given by~\eqref{r1} with $a=b-\beta/2$. 

We see that  due to~\eqref{ffp2} and  the multilinearity of a determinant,  
$G(x_1,\cdots,x_N;t)$~\eqref{fG} 
satisfies the diffusion equation.
\begin{align}
\frac{\partial}{\partial t}G(x_1,\cdots,x_N;t)
=\frac{1}{2}\sum_{j=1}^N\frac{\partial^2}{\partial x_j^2}G(x_1,\cdots,x_N;t).
\label{Rdiffusion}
\end{align}
In addition, by~\eqref{ffp4},  
it satisfies the condition 
\begin{align}
-\frac{\beta^2}{\pi^2}\int_{-\infty}^\infty dx_{j+1}\frac{e^{-\frac{\beta}{2}(x_{j+1}-x_j)}}
{e^{{\beta}(x_{j+1}-x_j)}-1}G(x_1,\cdots,x_N;t)=0.
\label{Rbc}
\end{align}
for $j=1,2,\cdots, N-1$. Though this condition is unusual, we will see that it is 
regarded as a finite temperature generalization of  the Neumann boundary
conditions at $x_j=x_{j+1},~j=1,\cdots, N-1$ in the zero temperature limit (see~\eqref{rbc}).

On the other hand, from~\eqref{ffp2} with the harmonicity of the Vandermonde 
determinant in~\eqref{w}, we see that $W(x_1,\cdots,x_N;t)$ 
satisfies
the Kolmogorov forward equation of the GUE Dyson's Brownian motion~\cite{Dys1962}, 
which is a dynamical generalization of the GUE,
\begin{multline}
\frac{\partial}{\partial t}W(x_1,\cdots,x_N;t)=\frac12\sum_{j=1}^N\frac{\partial^2}{\partial x_j^2}W(x_1,\cdots,x_N;t)\\-\sum_{ j=1}^N
\frac{\partial}{\partial x_j}\left(\sum_{\substack{m=1\\m\neq j}}^N\frac{1}{x_j-x_m}
\right)W(x_1,\cdots,x_N;t).
\label{dgueb}
\end{multline}
The time evolution equation for the GUE Dyson's Brownian motion can be transformed to
the imaginary-time Schr{\"o}dinger equation with free-Fermionic 
Hamiltonian (e.g. see Chapter 11 in~\cite{Fo2010}).
On the other hand note that the density function of the Whittaker measure~\eqref{whitm0} does not 
solve such a simple free-Fermionic time evolution equation~\eqref{dgueb}.

%

\subsection{The zero-temperature limit and a Brownian particle system with reflection 
interactions}\label{gtmarginals}
Let us consider the zero temperature limit of the equations~\eqref{Rdiffusion}
with~\eqref{Rbc} and~\eqref{dgueb}.
Note that for $x\neq 0$,
\begin{align}
-\lim_{\beta\rightarrow\infty}\tilde{f}_B(x)=1_{>0}(x),~-\lim_{\beta\rightarrow\infty}f_B(x)
=\lim_{\beta\rightarrow\infty}f_F(x)=1_{<0}(x),
\label{stepr}
\end{align}
where $\tilde{f}_B(x)$ is defined below~\eqref{th93} and $1_{>0}(x)$ and $1_{<0}(x)$ are the step functions defined by
\begin{align}
1_{>0}(x)=
\begin{cases}
1, & x >0,\\
0, & x\le 0,
\end{cases}
~~1_{<0}(x)=
\begin{cases}
0, & x>0,\\
1, & x\le 0.
\end{cases}
\label{step}
\end{align}
In addition we have
\begin{align}
\lim_{\beta\rightarrow\infty}F_{jk}(x;t)=\F_{j-k}(x;t),
\label{ffflim}
\end{align}
where $\F_n(x;t)$ is defined for $n\in\Z$ and $\e>0$ as
\begin{align}
\F_{n}(x;t)=\int_{i\R-\e}\frac{d\lambda}{2\pi i}
\frac{e^{-\lambda x+\lambda^2 t/2}}{\lambda^n}.
\label{cfdef}
\end{align}
Here we summarize a few properties 
of the function which are the zero temperature limit 
of~\eqref{fp1}--\eqref{fp4} for $F_{jk}(x;t)$.
\begin{gather}
\F_{-k}(x;t)=\lim_{\beta\rightarrow\infty}\psi_k(x;t)=
\frac{e^{-x^2/2t}}{\sqrt{2\pi t}}
\left(\frac{1}{2t}\right)^{\frac{k}{2}}H_k(x/\sqrt{2t}),~k=0,1,2,\cdots,
\label{fp1}\\
\frac{\partial}{\partial t}\F_n(x;t)=\frac{1}{2}\frac{\partial^2}{\partial x^2}\F_n(x;t),
\label{fp2}\\
\int_{-\infty}^{y}dx\F_n(x;t)=-\F_{n+1}(y;t),
\label{fp3}\\
\frac{\partial}{\partial x }\F_n(x;t)=-\F_{n-1}(y;t),
\label{fp4}
\end{gather}
where in~\eqref{fp1}, $\psi_k(x;t)$ is defined by~\eqref{psi} and 
$H_k(x)$ is the $k$th order Hermite polynomial~\cite{AAR1999}.
The second equality in~\eqref{fp1} has appeared as~\eqref{psibii}.
Note that~\eqref{fp4} corresponds to the zero-temperature 
limit of~\eqref{ffp4}, since RHS of~\eqref{ffp5} goes to
$-be^{-bx}$ in the zero-temperature limit and thus
the integral operator with the kernel
$\pi^2e^{\beta(y-x)/2}/\beta^2(e^{\beta(y-x)}-1)$ is equivalent 
to differentiation in the zero temperature 
limit when its action is restricted to $e^{-b x}$.

Let $\mathcal{G}(x_1,\cdots,x_N;t)$ be the zero-temperature limit of 
$G(x_1,\cdots,x_N;t)$~\eqref{fG} defined on $\R^N$.  From~\eqref{ffflim}, we find 
\begin{align}
\mathcal{G}(x_1,\cdots,x_N;t)=\det\left(\F_{j-k}(x_{N-j+1};t)\right)_{j,k=1}^N.
\label{Wf}
\end{align}
The function $\G(x_1,\cdots,x_N;t)$ appeared as a solution to
the Schr{\"o}dinger equation for the derivative nonlinear Schr{\"o}dinger 
type model~\cite{SaWa1998}. As discussed in~\cite{SaWa1998}, 
using~\eqref{fp2} and~\eqref{fp4} with basic properties of a determinant, 
we find that for $x_1\neq\cdots\neq x_N$,
$\G(x_1,\cdots,x_N;t)$ satisfies the diffusion equation,
\begin{align}
&\frac{\partial}{\partial t}\G(x_1,\cdots,x_N;t)
=\frac{1}{2}\sum_{j=1}^N\frac{\partial^2}{\partial x_j^2} \G(x_1,\cdots,x_N;t),
\label{rdiffusion}
\end{align}
with the boundary condition
\begin{align}
\frac{d}{dx_{j}}\G(x_1,\cdots,x_N;t)|_{x_{j}\rightarrow x_{j+1}}=0, {\text{~for~}}j=1,\cdots,N-1.
\label{rbc}
\end{align}

The probabilistic interpretation of $\G(x_1,\cdots,x_N;t)$ has been given in~\cite{Wa2007}.
Let $X_i(t),~i=1,\cdots,N$ be the stochastic processes with 
$N$-components  described by
\begin{align}
X_i(t)=y_i+B_i(t)+L^-_i(t),
\label{refBM}
\end{align}
where $y_i\in\R$ satisfying $y_1<y_2<\cdots <y_N$ represent 
initial positions, $B_i(t)$ denotes the standard 
Brownian motion and $L^-_i(t)$ is twice the semimartingale local 
time at zero of $X_i-X_{i-1}$ for $i=2,\cdots,N$ while $L^-_1(t)=0$.
The system~\eqref{refBM} describes the $N$-Brownian particles system with 
one-sided reflection interaction, i.e. the $i$th particle is reflected from the $i-1$th particle for
$i=2,3,\cdots,N$. 
In~\cite{Wa2007}, Warren found that the transition
density of this system from $y_i$ to $x_i$, $i=1,\cdots,N$
is written as $\G(x_1-y_1,\cdots,x_N-y_N;t)$. 
Such kind of  
determinantal transition density
was first obtained  for the 
totally asymmetric  simple exclusion process (TASEP)
in~\cite{Sc1997}. Furthermore,  based on the determinantal structures, 
various techniques for discussing the space-time joint distributions for the particle positions or
current have been developed for TASEP~\cite{NaSa2004,RaSc2005,Sa2005,BFPS2007,BFP2007,
BF2008, BFS2008(1), BFS2008(2),BFS2009} and the reflected Brownian particle 
system~\eqref{refBM}~\cite{FSW2015,FSW2015p}.

On the other hand, we have seen in~\eqref{lGUE} that the zero temperature limit of
$W(x_1,\cdots,x_N;t)$~\eqref{w} is the GUE density 
$P_{\text{GUE}}(x_1,\cdots,x_N;t)$~\eqref{t0result0}.
Note that $P_{\text{GUE}}(x_1,\cdots,x_N;t)$ also satisfies~ \eqref{dgueb}
since it holds for arbitrary $\beta$ i.e:
\begin{multline}
\frac{\partial}{\partial t}P_{\text{GUE}}(x_1,\cdots,x_N;t)=\frac12\sum_{j=1}^N\frac{\partial^2}{\partial x_j^2}P_{\text{GUE}}(x_1,\cdots,x_N;t)\\
-\sum_{ j=1}^N
\frac{\partial}{\partial x_j}\left(\sum_{\substack{m=1\\m\neq j}}^N\frac{1}{x_j-x_m}
\right)P_{\text{GUE}}(x_1,\cdots,x_N;t).
\label{dgue}
\end{multline}

From~\eqref{lGUE},~\eqref{stepr}, and~\eqref{ffflim},
we find that the zero-temperature limit of~\eqref{trelation} is
\begin{align}
\int_{(-\infty,u]^{N}}\prod_{\ell=1}^Ndx_{\ell}\cdot\mathcal{G}(x_1,\cdots,x_N;t)
=
\int_{(-\infty,u]^{N}}\prod_{j=1}^Ndx_{j}\cdot P_{\text{GUE}}(x_1,\cdots,x_N;t).
\label{wresult}
\end{align}
In \cite{Wa2007} Warren showed that this relation, which connects the
two different processes, is obtained in the following way. First  one introduces 
a process on the $N(N+1)/2$-dimensional Gelfand-Tsetlin cone
whose two marginals describe the above two processes.
The Gelfand-Tsetlin cone 
$\text{GT}_k,~k=1,2,\cdots$ is defined as
\begin{align}
\text{GT}_k:=\{(x^{(1)},\cdots,x^{(k)})|~&x^{(i)}=(x^{(i)}_1,\cdots,x^{(i)}_i)\in\R^i
\text{~with~} i=1,\cdots,k, \notag\\
&x^{(m+1)}_{\ell+1}\le x^{(m)}_{\ell}
\le x^{(m+1)}_{\ell} \text{~with~}1\le \ell\le m\le k-1\}.
\label{gtn}
\end{align}
For the graphical representation of an element of $\text{GT}_k$, see Fig.~\ref{vtgt} (b).
Next we introduce a following stochastic process 
on $\text{GT}_N$.
Let $(X^{(1)}(t),\cdots,X^{(N)}(t))$ with $X^{(j)}(t)=(X^{(j)}_1(t),\cdots,X^{(j)}_j(t))$ be
a process defined by
\begin{align}
X^{(j)}_i(t)=B^{(j)}_i(t)+y^{(j)}_i+L^{(j)-}_i(t)-L^{(j)+}_i(t),~1\le i\le j\le N,
\label{ws}
\end{align}
where $B^{(j)}_i(t)$ are the $N(N+1)/2$ independent Brownian motions
starting at the origin,
$y^{(j)}_i$ represent the initial positions and 
the process $L^{(j)-}_i(t)$ and $L^{(j)+}_i(t)$ are twice the semimartingale
local time at zero of $X^{(j)}_i-X^{(j-1)}_{i}$ and $X^{(j)}_i-X^{(j-1)}_{i-1}$ respectively.
Eq.~\eqref{ws}  describes the interacting particle systems where each 
$X^{(j)}_i(t)$ is a Brownian motion reflected
from $X^{(j-1)}_{i-1}(t)$ to a negative direction and from $X^{(j-1)}_i(t)$ to a positive direction.
In~\cite{BoFe2014}, Borodin and Ferrari also introduced 
similar processes on the discrete Gelfand-Tsetlin cone
where the 
probability measure at a particular time is 
described by the Schur process~\cite{OkRe2003}.

The pdf of the system~\eqref{ws} at time $t$ can be given explicitly : 
For the case of $y^{(j)}_i=0$, it is expressed as
\begin{align}
\mathcal{Q}_{\rm GT}(\underline{x}_N;t)=\prod_{1\le i<j\le N}
\left(x^{(N)}_i-x^{(N)}_j\right)\cdot\prod_{k=1}^N 
\frac{\exp{\left(-\left(x^{(N)}_k\right)^2/2t\right)}}{t^{k-1}\sqrt{2\pi t}}\cdot
1_{\text{GT}}(\underline{x}_N).
\label{Wmar}
\end{align}
where $\underline{x}_N$ is defined above~\eqref{taweight} and 
$1_{\text{GT}}(\underline{x}_k)$ represents the indicator function on GT$_k$.
The pdfs of the 
two marginals, $(x^{(1)}_1,\cdots,x^{(N)}_1)$ and $(x^{(N)}_1,\cdots,x^{(N)}_N)$
for $\mathcal{Q}_{\text{GT}}(\underline{x}_N;t)$ 
was obtained as follows:

\begin{proposition}(Proposition 6 and 8 in~\cite{Wa2007})\label{propw}
\begin{align}
&\int_{\R^{N(N-1)/2}}dA_1\mathcal{Q}_{\rm GT}(\underline{x}_N;t)
=\G(x^{(1)}_1,\cdots,x^{(N)}_1;t)\prod_{j=1}^{N-1}1_{>0}(x^{(j+1)}_1-x^{(j)}_1),
\label{dA0}
\\
&\int_{\R^{N(N-1)/2}}dA_2\mathcal{Q}_{\rm GT}(\underline{x}_N;t)=
N!P_{\rm{GUE}}(x^{(N)}_1,\cdots,x^{(N)}_N;t)\prod_{j=1}^{N-1}1_{>0}(x^{(N)}_j-x^{(N)}_{j+1}),
\label{dB0}
\end{align}
where $\G(x_1,\cdots,x_N;t)$, $P_{\rm{GUE}}(x^{(N)}_1,\cdots,x^{(N)}_N;t)$,~$1_{>0}(x)$ 
and $dA_1,~dA_2$ are defined 
by~\eqref{Wf}, \eqref{t0result0}, \eqref{step}  and~\eqref{dAdB} respectively.
\end{proposition}
\noindent
{\bf Remark.}
Note that $\G(x^{(1)}_1,\cdots,x^{(N)}_1;t)$ in~\eqref{dA0} can be replaced 
by an arbitrary function on $\R^N$such that it corresponds to  
$\G(x^{(1)}_1,\cdots,x^{(N)}_1;t)$ in the region $x^{(1)}_1<x^{(2)}_1<\cdots<x^{(N)}_1$.
For later discussion on a generalization of finite temperature, we chose it as
$\G(x^{(1)}_1,\cdots,x^{(N)}_1;t)$ on the whole $\R^N$.

We see that the relation~\eqref{wresult} is obtained from this theorem.
By decomposing the integral on $\underline{x}_N$ in two 
different ways,  we clearly have
\begin{align}
\int_{(-\infty,u]^{N(N-1)/2}}d\underline{x}_N\mathcal{Q}_{\text{GT}}(\underline{x}_N;t)
&=\int_{(-\infty,u]^N}\prod_{j=1}^{N}dx^{(j)}_1
\int_{\R^{N(N-1)/2}}dA_1\mathcal{Q}_{\rm GT}(\underline{x}_N;t)\notag\\
&=\int_{(-\infty,u]^N}\prod_{j=1}^{N}dx^{(N)}_{j}
\int_{\R^{N(N-1)/2}}dA_2\mathcal{Q}_{\rm GT}(\underline{x}_N;t).
\label{GT1}
\end{align}
Applying~\eqref{dA0} and~\eqref{dB0} to this equation, we get
\begin{multline}
\int_{(-\infty,u]^N}\prod_{j=1}^{N}dx^{(j)}_1
\G(x^{(1)}_1,\cdots,x^{(N)}_1;t)\prod_{j=1}^{N-1}1_{>0}(x^{(j+1)}_1-x^{(j)}_1)\\
=\int_{(-\infty,u]^N}\prod_{j=1}^{N}dx^{(N)}_{j}
N!P_{\rm{GUE}}(x^{(N)}_1,\cdots,x^{(N)}_N;t)\prod_{j=1}^{N-1}1_{>0}(x^{(N)}_j-x^{(N)}_{j+1}).
\label{GT3}
\end{multline}
Due to the symmetry of $P_{\text{GUE}}(x_1,\cdots,x_N;t)$ under the permutations
of $x_1,\cdots,x_N$, we readily see that RHS of this equation is equal 
to RHS of~\eqref{wresult}. Also we find that LHS of~\eqref{GT3} becomes
\begin{align}
&~\int_{(-\infty,u]^N}\prod_{j=1}^{N}dx^{(j)}_1\cdot
\G(x^{(1)}_1,\cdots,x^{(N)}_1;t)\prod_{j=1}^{N-1}1_{>0}(x^{(j+1)}_1-x^{(j)}_1)\notag\\
&=\int_{(-\infty,u]^N}\prod_{j=1}^{N}dx^{(j)}_1\cdot
\G(x^{(1)}_1,\cdots,x^{(N)}_1;t)\prod_{j=1}^{N-1}\left(1_{>0}(x^{(j+1)}_1-x^{(j)}_1)+1_{>0}(x^{(j)}_1-x^{(j+1)}_1)\right)\notag\\
&=\int_{(-\infty,u]^N}\prod_{j=1}^{N}dx^{(j)}_1\cdot
\G(x^{(1)}_1,\cdots,x^{(N)}_1;t),
\end{align}
where in the first equality we used for $k=2,\cdots, N$ and 
$(x^{(1)}_1,\cdots,x^{(N)}_1)\in (-\infty,u]^N$ 
\begin{align}
\int_{(-\infty,u]}dx^{(k)}_1
\G(x^{(1)}_1,\cdots,x^{(N)}_1;t)\prod_{j=1}^{k-2}1_{>0}(x^{(j+1)}_1-x_1^{(j)})\cdot
1_{>0}(x^{(k-1)}_1-x^{(k)}_1)=0.
\label{GT2}
\end{align}
Note that $\G(x^{(1)}_1,\cdots,x^{(N)}_1;t)$ is defined on $\R^N$ and is finite
even outside the region $x^{(1)}_1<x^{(2)}_1<\cdots<x^{(N)}_1$. (See Remark.~of 
Proposition~\ref{propw}.)
Eq.~\eqref{GT2} is obtained from the following observation: 
putting the last factor $1_{>0}(x^{(k-1)}_1-x^{(k)}_1)$  in the $N-k-1$th row of the
determinant  $\G(x^{(1)}_1,\cdots,x^{(N)}_1;t)$ in~\eqref{GT2} then applying~\eqref{fp3}, 
we get  the determinant which has the same two rows. 

Thus~\eqref{wresult} is obtained from Proposition~\ref{propw}.
This is similar to the situation of~\eqref{trelation} and Theorem~\ref{mar}.
This naive observation gives us the impression that the pdf
$\mathcal{Q}_{\text{GT}}(\underline{x}_N;t)$~\eqref{Wmar} is the zero-temperature limit of
the weight $R_u(\underline{x}_N;t)$~\eqref{taweight}.  
However in fact this is not the case. Let $\mathcal{R}_u(\underline{x}_N;t):=\lim_{\beta\rightarrow
\infty}R_u(\underline{x}_N;t)$. From~\eqref{stepr} and \eqref{ffflim} one has
\begin{align}
\mathcal{R}_u(\underline{x}_N;t)=(-1)^{N(N-1)/2}\det\left(\F_{1-i}(x^{(N)}_j;t)\right)_{i,j=1}^N 
\prod_{1\le j\le k\le N}1_{>0}(x^{(k-1)}_{j-1}-x^{(k)}_{j}).
\label{wv2}
\end{align}
From~\eqref{dpsilim} and~\eqref{fp1}, it is further rewritten as
\begin{align}
\mathcal{R}_u(\underline{x}_N;t)=
\prod_{1\le i<j\le N}
\left(x^{(N)}_i-x^{(N)}_j\right)\cdot\prod_{k=1}^N 
\frac{\exp{\left(-\left({x^{(N)}_k}\right)^2/{2t}\right)}}{t^{k-1}\sqrt{2\pi t}}
1_{>0}\left(u-x^{(k)}_1\right)
\cdot
1_{V_N}(\underline{x}_N),
\label{RR}
\end{align}
where $1_{V_k}(\underline{x}_k)$ is the indicator function on
an ordered set $V_k$ defined by
\begin{align}
V_k:=\{(x^{(1)},\cdots,x^{(k)})|~x^{(j)}=(x^{(j)}_1,\cdots,x^{(j)}_j)\in\R^j,
x^{(m+1)}_{\ell+1}
\le x^{(m)}_{\ell},~1\le \ell\le m\le k-1\}.
\label{vn}
\end{align}
For the graphical representation of an element of~\eqref{vn}, see Fig.~\ref{vtgt} (a).
Comparing~\eqref{Wmar} with~\eqref{RR}, we see that they have the same form
but their supports ($\text{GT}_N$ and $V_N$) are different.
We further notice that
$V_N$ with an additional order $x^{(m)}_{\ell}\le x^{(m+1)}_{\ell},~1\le \ell\le m\le N-1$ 
corresponds to GT$_N$. 

Hence our approach using $\mathcal{R}_u(\underline{x}_N;t)$
can be regarded as a modification of Warren's arguments on~$\text{GT}_N$
to the ones on the partially ordered spece $V_N$.
Let us focus on two marginals $(x^{(1)}_1,x^{(2)}_1,\cdots, x^{(N)}_1)$ and $(x^{(N)}_1,x^{(N)}_2,
\cdots, x^{(N)}_N)$ for $\mathcal{R}_u(\underline{x}_N;t)$~\eqref{RR}.
By taking the zero-temperature limit of Theorem~\ref{mar}, we have 
the following analogue of Proposition~\ref{propw}: 
\begin{proposition}\label{mar0}
\begin{align}
&\int_{\R^{N(N-1)/2}}dA_1\,\mathcal{R}_u(\underline{x}_N;t)
=\mathcal{G}(x^{(1)}_1,\cdots,x^{(N)}_1;t)\prod_{j=1}^N1_{>0}(u-x^{(j)}_1),
\label{dA}\\
&~\int_{\R^{N(N-1)/2}}d\,A_2~\mathcal{R}_{u}(\underline{x}_N;t)=
P_u\left(x^{(N)}_1,\cdots,x^{(N)}_N;t\right)\prod_{j=1}^N1_{>0}\left(u-x^{(N)}_{j}\right),
\label{dB}
\end{align}
where for the definition of $dA_1$ and $dA_2$, see~\eqref{dAdB}, 
$\mathcal{G}(x^{(1)}_1,\cdots,x^{(N)}_1;t)$ is given by~\eqref{Wf} and
\begin{align}
P_u\left(x^{(N)}_1,\cdots,x^{(N)}_N;t\right)
=\prod_{j=1}^{N}
\frac{(u-x^{(N)}_{j})^{j-1}}{(j-1)!t^{j-1}}\cdot\prod_{1\le j<k\le N}\left(x^{(N)}_j-x^{(N)}_k\right)\cdot\prod_{j=1}^N
\frac{e^{-\left(x^{(N)}_j\right)^2/2t}}{\sqrt{2\pi t}}.
\label{pu}
\end{align} 
\end{proposition}

\smallskip
\noindent
{\bf Proof.} It is obtained by taking the zero-temperature limit $(\beta\rightarrow\infty)$ 
in Theorem~\ref{mar}.
%
%
\qed
\smallskip

As discussed in~\eqref{align}, $P_{\text{GUE}}(x_1,\cdots,x_N;t)$~\eqref{t0result0}  can be interpreted as 
the symmetric version of $P_u(x_1,\cdots,x_N;t)$:
\begin{align}
\frac{1}{N!}\sum_{\sigma^{(N)}\in S_N}P_u(x_{\sigma^{(N)}(1)},\cdots,x_{\sigma^{(N)}(N)};t)
=P_{\text{GUE}}(x^{(N)}_1,\cdots,x^{(N)}_N;t).
\label{gsym2} 
\end{align}
Therefore by the similar discussion in~\eqref{GT1}, we see that the relation~\eqref{wresult} 
is obtained also from Proposition~\ref{mar0}. 

The fact that both 
Proposition~\ref{propw} and~\ref{mar0} lead to~\eqref{wresult} implies the relation
\begin{align}
\int_{\R^{N(N+1)/2}} d\underline{x}_N
\mathcal{R}_{u}(\underline{x}_N;t)
=
\int_{(-\infty,u]^{N(N+1)/2}}d\underline{x}_N
\mathcal{Q}_{\rm GT}(\underline{x}_N;t).
\label{p11}
\end{align}
This equivalence of their integration values 
is generalized in the following way.
%
%
\smallskip
\begin{proposition}\label{vgt}
Let $f_{\rm sym}(\underline{x}_N) $ be the function defined above~\eqref{rrbar}. Then 
we have
\begin{align}
\int_{\R^{N(N+1)/2}} d\underline{x}_N
f_{\rm {sym}}(\underline{x}_N)\mathcal{R}_{u}(\underline{x}_N;t)
=
\int_{(-\infty,u]^{N(N+1)/2}}d\underline{x}_N
f_{\rm{sym}}(\underline{x}_N)\mathcal{Q}_{\rm GT}(\underline{x}_N;t)
\label{p11}
\end{align}
\end{proposition}
\smallskip
An essential step of the proof of this proposition is represented as the following
\begin{lemma}\label{vgt2}
\begin{align}
\sum_{\sigma^{(j)}\in S_j,j=1,\cdots,N}\sgn\sigma^{(N)}1_{V_N}(\underline{x}_N^\sigma)
=\sum_{\sigma^{(j)}\in S_j,j=1,\cdots,N}\sgn\sigma^{(N)}1_{\rm{GT}}(\underline{x}_N^\sigma)
\label{evgt}
\end{align}
\end{lemma}
\smallskip
The proof of this lemma will be given in Appendix~\ref{pvgt}.
Using this lemma we readily derive Proposition~\ref{vgt}.

\noindent
{\bf Proof of Proposition~\ref{vgt}.}
Substituting the definition of $\mathcal{R}_u(\underline{x}_N;t)$~\eqref{RR} 
into~\eqref{p11}, we see that the LHS of~\eqref{p11} is rewritten as
\begin{align}
&~~\int_{\R^{N(N+1)/2}}d\underline{x}_Nf_{\text{sym}}(\underline{x}_N)
\prod_{k=1}^N1_{>0}(u-x^{(k)}_1)e^{-\left(x^{(N)}_k\right)^2/2t}\cdot
\prod_{1\le i<j\le N}\left(x^{(N)}_i-x^{(N)}_j\right)\notag\\
&\hspace{10cm}\times\sum_{\sigma^{(j)}\in S_j,j=1,\cdots,N}
\sgn\sigma^{(N)}1_V(\underline{x}_N^{\sigma})\notag\\
&=
\int_{\R^{N(N+1)/2}}d\underline{x}_Nf_{\text{sym}}(\underline{x}_N)
\prod_{k=1}^N1_{>0}(u-x^{(k)}_1)e^{-\left(x^{(N)}_k\right)^2/2t}\cdot
\prod_{1\le i<j\le N}\left(x^{(N)}_i-x^{(N)}_j\right)\notag\\
&\hspace{10cm}\times\sum_{\sigma^{(j)}\in S_j,j=1,\cdots,N}
\sgn\sigma^{(N)}1_{\text{GT}}(\underline{x}_N^{\sigma})\notag\\
&=\int_{\R^{N(N+1)/2}}d\underline{x}_Nf_{\text{sym}}(\underline{x}_N)\prod_{k=1}^N1_{>0}(u-x^{(k)}_1)
\sum_{\sigma^{(j)}\in S_j,j=1,\cdots,N}
\mathcal{Q}_{\text{GT}}(\underline{x}_N^{\sigma};t)
\end{align}
where in the second equality we use Lemma~\ref{vgt2}.
\qed
\smallskip

\section{Fredholm determinant formulas}\label{fdf}
\subsection{A Fredholm determinant with a biorthogonal kernel}
The function $W(x_1,\cdots,x_N;t)$~\eqref{w1} has a notable determinantal 
structure that it is described by a product of two 
determinants. This allows us to apply the results of random matrix theory and determinantal point 
processes developed in~\cite{TW1998,Jo2003} and to get the Fredholm determinant
representation. 

To see this we provide a lemma.  Let
$\phi_j(x;t),~j=0,1,2,\cdots$ be
\begin{align}
\phi_j(x;t)=\frac{1}{2\pi i}\oint dv\, e^{vx-v^2t/2}\frac{\Gamma(1+v/\beta)^N}
{v^{j+1}},
\label{phi}
\end{align}
where the contour encloses the origin anticlockwise with radius smaller 
than $\beta$. 
We find $\phi_j(x;t)$ and $\psi_k(x;t)$~\eqref{psi}
satisfy the biorthonormal relation:
\begin{lemma}\label{lembior}
For $j,k\in \{0,1,2,\cdots\}$, we have
\begin{align}
\int_{-\infty}^{\infty}dx\,\phi_j(x;t)\psi_k(x;t)=\delta_{j,k}.
\label{bior}
\end{align}
\end{lemma}

\noindent
{\bf Proof.}
Substituting the definitions~\eqref{psi} and~\eqref{phi}
into LHS of~\eqref{bior}, one has
\begin{align}
&~~\int_{-\infty}^{\infty}dx\,\phi_j(x;t)\psi_k(x;t)\notag\\
&=\frac{1}{(2\pi)^2i}
\oint dv\int_{-\infty}^{\infty}dw\, e^{-(w^2+v^2)t/2}
\left(\frac{\Gamma(1+v/\beta)}{\Gamma(1+iw/\beta)}\right)^N
\frac{(iw)^k}{v^{j+1}}
\int_{-\infty}^{\infty}dx\,e^{(v-iw)x}.
\end{align}
As the integrand in this equation is analytic on $\C$ 
with respect to $w$, we can shift the integration path 
as $w=w'-i v,~w'\in\R$.
Then using 
\begin{align}
\frac{1}{2\pi}\int_{-\infty}^{\infty} dx\, e^{(v-iw)x}=
\frac{1}{2\pi}\int_{-\infty}^{\infty} dx\, e^{-iw'x}=
\delta(w'),
\end{align}
we find
\begin{align}
\int_{-\infty}^{\infty}dx\,\phi_j(x;t)\psi_k(x;t)
=\frac{1}{2\pi i}\oint dv\,v^{k-j-1}=\delta_{j,k}.
\end{align}
\qed

\smallskip
The residue calculus shows that the function $\phi_j(x;t)$ is a $j$th order polynomial 
in $x$ and the coefficient of the highest order is $1/j!$. 
As the Vandermonde determinant in~\eqref{w} 
is expressed as 
\begin{align}
\prod_{1\le j<k\le N}(x_k-x_j)=\det\left(x^{j-1}_k\right)_{j,k=1}^N
=\det\left((j-1)!\phi_{j-1}(x_k,t)\right)_{j,k=1}^N,
\end{align} 
$W(x_1,\cdots,x_N;t)$ is rewritten as a product of two determinants
\begin{align}
W(x_1,\cdots,x_N;t)=\frac{1}{N!}\det\left(\phi_{j-1}(x_k;t)\right)_{j,k=1}^N
\det\left(\psi_{j-1}(x_k;t)\right)_{j,k=1}^N.
\label{detpp}
\end{align}
From Lemma~\ref{lembior} and~\eqref{detpp}, we obtain a Fredholm determinant representation
for the moment generating function. Throughout this paper, we follow~\cite{BC2014} for the notation on Fredholm determinants. 
\begin{proposition}\label{propfred1}
\begin{align}
\int_{-\infty}^{\infty}\prod_{j=1}^Ndx_j\, g(x_j)\cdot W(x_1,\cdots,x_N;t)
=\det\left(1-\bar{g}K\right)_{L^2(\R)}
\label{propfred}
\end{align}
where $g(x)$ is an arbitrary function such that the left hand side
is well-defined and 
in the right hand side $\det\left(1-\bar{g}K\right)_{L^2(\R)}$ 
represents 
a Fredholm determinant
defined by
\begin{align}
\det\left(1-\bar{g}K\right)_{L^2(\R)}=\sum_{k=0}^{\infty}\frac{(-1)^k}{k!}\int_{\R^k}
\prod_{j=1}^k dx_j\, \bar{g}(x_j)\cdot\det\left(K(x_l,x_m;t)
\right)_{l,m=1}^k.
\label{Fred}
\end{align}
Here $\bar{g}(x)=1-g(x)$ and $K(x,y;t)$ is written in terms of
the biorthogonal functions $\psi_j(x,t)$~\eqref{psi} 
and $\phi_k(x,t)$~\eqref{phi} as
\begin{align}
&K(x,y;t)=\sum_{k=0}^{N-1}\phi_k(x;t)\psi_k(y;t).
\label{kernel}
\end{align}
\end{proposition}

\smallskip
\noindent
{\bf Proof.}
We readily obtain this representation  
by applying the techniques in~\cite{TW1998} with Lemma~\ref{lembior}
to~LHS of~\eqref{propfred}. For reference, here is an outline of the proof.
First,  using the Andr{\'e}ief (Cauchy-Binet) identity~\eqref{heine},
we have
\begin{align}
&~\int_{\R^N}\prod_{j=1}^Ndx_j\, g(x_j)\cdot W(x_1,\cdots,x_N;t)
=\det\left(\int_\R dx\, g(x)\phi_{j-1}(x;t)\psi_{k-1}(x;t)\right)_{j,k=1}^N\notag\\
&=\det\left(\int_\R dx\,\phi_{j-1}(x;t)\psi_{k-1}(x;t)- \int_\R dx\,\bar{g}(x)\phi_{j-1}(x;t)\psi_{k-1}(x;t)\right)_{j,k=1}^N 
\notag\\
&=\det\left(\delta_{j,k}-A_{j,k}\right)_{j,k=1}^N,
\label{arep}
\end{align}
where $A_{j,k},~j,k=1,\cdots,N$ is defined as
\begin{align}
A_{jk}=\int_\R dx\, \bar{g}(x)\phi_{j-1}(x;t)\psi_{k-1}(x;t).
\end{align}
In the first equality of~\eqref{arep}, we used~\eqref{detpp} with~\eqref{heine} and in the last one
we used Lemma~\ref{lembior}. We further rewrite $A_{jk}$ as
\begin{align}
A_{jk}=\int_{\R}dx\, B(j,x)C(x,k)
\end{align}
by using
\begin{align}
B(j,x)=\phi_{j-1}(x;t), ~C(x,k)=\bar{g}(x)\psi_{k-1}(x;t).
\end{align}
Applying the identity for Fredholm determinants, 
\begin{align}
\det\left(\delta_{j,k}-A_{j,k}\right)_{j,k=1}^N=\det(1-BC)_{L^2(\{1,2,\cdots,N\})}=\det(1-CB)_{L^2(\R)},
\end{align}
and noting
\begin{align}
(CB)(x,y)=\bar{g}(x)\sum_{k=0}^{N-1}\phi_{k}(x;t)\psi_k(y;t),
\end{align}
we arrive at our desired expression.
\qed

Combining this proposition with Theorem~\ref{thmmain},
we readily obtain
\begin{corollary}\label{corfred}
\begin{align}
\E\left(e^{-\frac{e^{-\beta u} Z_N(t)}{\beta^{2(N-1)}}}\right)
=\det\left(1-\bar{f}_uK\right)_{L^2(\R)}
\label{cor4}
\end{align}
where the right hand side is the Fredholm determinant~\eqref{Fred}
with the kernel $\bar{f}_u(x_i)K(x_i,x_j;t)$,
$\bar{f}_u(x_j)=1-f_F(x_j-u)$,
and $K(x_i,x_j;t)$ is defined in~\eqref{kernel}.
\end{corollary}

\noindent
\smallskip
\noindent
As in~\eqref{psibii}, we see
\begin{align}
\lim_{\beta\rightarrow\infty}\phi_k(x;t)=
\frac{1}{2\pi i}\oint dv\, \frac{e^{vx-v^2t/2}}{
{v^{k+1}}}=\frac{1}{k!}\left(\frac{t}{2}\right)^{\frac{k}{2}}H_k\left(\frac{x}{\sqrt{2t}}\right),
\label{phibii}
\end{align}
which is due to another representation of the $n$th order Hermite polynomial $H_n(x)$
(see e.g. Section 6.1 in~\cite{AAR1999}),
\begin{align}
H_n(x)=\frac{n!}{2\pi i}\oint dz\, \frac{e^{2xz-z^2}}{z^{n+1}},
\label{hermite2}
\end{align}
where the contour encloses the origin anticlockwise.
 From~\eqref{psibii} and~\eqref{phibii}, we find
\begin{align}
\lim_{\beta\rightarrow\infty}K(x_1,x_2;t)
=\frac{e^{-x_2^2/2t}}{\sqrt{2\pi t}}\sum_{k=0}^{N-1}\frac{H_k(x_1/\sqrt{2t}) H_k(x_2/\sqrt{2t})}{2^kk!}.
\label{khermite}
\end{align}
Here RHS appears as a correlation kernel of the eigenvalues in 
the GUE random matrices~\cite{M2004}.

Thus $K(x_i,x_j;t)$ is a simple biorthogonal deformation of
the kernel with Hermite polynomials which appears in the eigenvalue correlations 
of $N\times N$ GUE random matrices. Using this Fredholm determinant
expression~\eqref{cor4}, we can understand a few asymptotic properties of the partition 
function by applying saddle point analyses to the kernel as will be discussed
in~Sec.~\ref{KPZsl}.

\subsection{A representation from the Macdonald processes}
In~\cite{O2012}, O'Connell first introduced the probability
measure on $\R^N$ which is called the Whittaker measure
$m(x_1,\cdots,x_N;t)\prod_{j=1}^Ndx_j$ whose 
density function $m(x_1,\cdots,x_N;t)$  is defined 
in terms of the Whittaker function $\Psi_{\lambda}(x_1,\cdots,x_N)$ (see~\cite{O2012}),
\begin{align}
m_t(x_1,\cdots,x_N;t)=\Psi_0(\beta x_1,\cdots,\beta x_N)
\int_{(i\R)^N}d\lambda\,
\Psi_{-\lambda/\beta}(\beta x_1,\cdots,\beta x_N)e^{\sum_{j=1}^N\lambda_j^2t/2}s_N(\lambda/\beta),
\label{whitm}
\end{align}
where throughout this paper we denote $\lambda=(\lambda_1,\cdots,\lambda_N)$  and 
$s_N(\lambda)$ is defined by~\eqref{zsk}.
Then he showed the following relation about the distribution of the
free energy
$F_N(t)=-\log(Z_N(t))/\beta$ 
(see Theorem 3.1 and
Corollary 4.1 in~\cite{O2012}),
\begin{align}
\text{Prob}\left(-F_N(t)+\frac{N-1}{\beta}\log\beta^2\le s\right)=
\int_{(-\infty, s]}dx_1\int_{\R^{N-1}}\prod_{j=2}^{N}
dx_{j}\cdot m(x_1,\cdots,x_N;t).
\label{whit}
\end{align}
The density function $m(x_1, \cdots, x_N;t)$~\eqref{whitm} is also a finite temperature 
extension of $P_{\text{GUE}}(x_1, \allowbreak\cdots, x_N;t)$ \eqref{t0result0}. 
Actually it has been known that 
$m(x_1,\cdots,x_N;t)$ converges to $P_{\text{GUE}}(x_1,\cdots\allowbreak,x_N;t)$ in the zero-temperature limit. 
(See Sec.6 in~\cite{O2012}). In contrast to $W(x_1,\cdots, x_N;t)$~\eqref{w}, however, this extension
does not inherit the determinantal structure which
$P_{\text{GUE}}(x_1,\cdots,x_N;t)$ has and thus we cannot apply the techniques in 
random matrix theory which is useful especially for asymptotic
analyses of the GUE. This fact necessitated the developments of
new methods~\cite{BC2014,BCF2012,BCS2012,
BCR2013,SS2010a,SS2010b,SS2010c,SS2010d,ACQ2010,
CLDR2010,D2010,D2010p2}. By using the techniques of 
the Macdonald difference operators~\cite{BC2014} and 
the duality~\cite{BCS2012}, one can get
a Fredholm determinant expression 
for the moment generating function of the partition function, 
which allows us to access the asymptotic properties.

\begin{proposition}(\cite{BC2014})
\begin{align}
\E\left(e^{-\frac{e^{-\beta u} Z_N(t)}{\beta^{2(N-1)}}}\right)
=\det\left(1+L\right)_{L^2(C_0)}
\label{bcrep}
\end{align}
where $C_0$ denotes the contour enclosing only the origin positively with radius
$r<\beta/2$
and the kernel $L(v,v';t)$ is written as
\begin{align}
L(v,v';t)=\frac{1}{2\pi i}\int_{i\R+\delta}dw\,\frac{\pi/\beta}{\sin(v'-w)/\beta}
\frac{w^Ne^{w^2t/2-wu}}{v'^Ne^{v'^2t/2-v'u}}\frac{1}{w-v}\frac{\Gamma(1+v'/\beta)^N}
{\Gamma(1+w/\beta)^N}.
\label{Lkernel}
\end{align}
Here $\delta$ satisfies the condition $r<\delta<\beta-r$.
\end{proposition}

We can show the equivalence between the two expressions~\eqref{cor4} and~\eqref{bcrep}.
\begin{proposition}\label{propfred2}
\begin{align}
\det(1-\bar{f}_uK)_{L^2(\R)}=\det(1+L)_{L^2(C_0)}
\label{prop5}
\end{align}
where $\bar{f}_u(x)=1-f_F(x-u)$ and $K(x,x';t)$
and $L(v,v';t)$ are defined~\eqref{kernel} 
and~\eqref{Lkernel}
respectively.
\end{proposition}

\noindent
{\bf Proof.}

Substituting the definitions~\eqref{psi} and~\eqref{phi} into~\eqref{kernel}, we have
\begin{align}
K(x,x';t)
&=\oint_{C_0} dv\int_{i\R+\delta}dw\, e^{vx-wx'-(v^2-w^2)t/2}
\frac{\Gamma(1+v/\beta)^N}{\Gamma(1+w/\beta)^N}
\frac{1}{v}\sum_{k=0}^{N-1}\left(\frac{w}{v}\right)^k\notag\\
&=\oint_{C_0} dv\int_{i\R+\delta}
dw\, e^{vx-wx'-(v^2-w^2)t/2}
\frac{\Gamma(1+v/\beta)^N}{\Gamma(1+w/\beta)^N}
\frac{1-(w/v)^N}{v-w}.
\label{v-w}
\end{align}
For the definition of $C_0$, see below~\eqref{bcrep}.
Here we changed 
$w\rightarrow -iw$ in~\eqref{psi} 
and shift the path of $w$ by $\delta$ which is 
larger than the radius of $v$. 
We notice that 
although the last expression in~\eqref{v-w}
consists of two terms proportional to $1/(v-w)$
 and $(w/v)^N/(v-w)$, 
the integration of the term proportional 
to $1/(v-w)$ with respect to $v$ vanishes.
Thus we see
\begin{align}
\bar{f}_u(x)K(x,x')&=-\bar{f}_u(x)\oint_{C_{0}} dv\int_{i\R+\delta}
dw\,\frac{e^{vx-wx'-(v^2-w^2)t/2}}{w-v}
\left(\frac{\Gamma(1+v/\beta)}{\Gamma(1+w/\beta)}
\frac{w}{v}\right)^N\notag\\
&=-\oint_{C_{0}}dv\, A(x,v)B(v,x')
\end{align}
where we set
\begin{align}
&A(x,v)=\bar{f}_u(x)e^{vx-v^2t/2}\left(\frac{\Gamma(1+v/\beta)}{v}\right)^N,\\
&B(v,x')=\int_{i\R+\delta}dw\,\frac{e^{-wx'+w^2t/2}}{w-v}
\left(\frac{w}{\Gamma(1+w/\beta)}\right)^N.
\end{align}
Here we use the relation for Fredholm determinants, 
$\det(1-AB)_{L^2(\R)}=\det(1-BA)_{L^2(C_0)}$, where the kernel $-(BA)(v,v')$
on RHS reads
\begin{align}
&~~-\int_{-\infty}^\infty dx\, B(v,x)A(x,v')
\notag
\\
&=-\int_{i\R+\delta}dw\,\frac{e^{(w^2-v'^2)t/2}}{w-v}
\left(\frac{w\Gamma(1+v'/\beta)}{v'\Gamma(1+w/\beta)}\right)^N
\int_{-\infty}^{\infty}dx\, \bar{f}_u(x)e^{(v'-w)x}.
\label{Lrelation}
\end{align}
Using the relation 
\begin{align}
\int_{-\infty}^{\infty}dx\,\frac{e^{ax}}{1+e^{x}}=\frac{\pi}{\sin\pi a},~~
\text{for~}0<\text{Re}~a<1,
\label{sinrelation}
\end{align}
we perform the integration over $x$ in~\eqref{Lrelation} as
\begin{align}
-~\int_{-\infty}^{\infty}dx\, \bar{f}_u(x)e^{(v'-w)x}
=\int_{-\infty}^{\infty}dx\,
\frac{-e^{\beta(x-u)+(v'-w)x}}{1+e^{\beta(x-u)}}
=\frac{e^{(v'-w)u}\pi/\beta}{\sin\left[(v'-w)\pi/\beta\right]}.
\label{xint}
\end{align}

Note that because of the conditions $0<r<\beta/2$ and 
$r<\delta<\beta-r$ (see below~\eqref{bcrep} and~\eqref{Lkernel} respectively),
\eqref{sinrelation} is applicable to the above equation.
Thus from~\eqref{Lrelation} and~\eqref{xint}, we have
\begin{align}
-\int_{-\infty}^\infty dx\, B(v,x)A(x,v')&=
\frac{1}{2\pi i}\int_{i\R+\delta}dw\,\frac{\pi/\beta}{\sin\left[(v'-w)\pi/\beta\right]}
\frac{w^Ne^{w^2t/2-wu}}{v'^Ne^{v'^2t/2-v'u}}\frac{1}{w-v}\frac{\Gamma(1+v'/\beta)^N}
{\Gamma(1+w/\beta)^N}\notag\\
&=L(v,v';t)
\end{align}
\qed

\section{The scaling limit to the KPZ equation}\label{KPZsl}
In this section,  we discuss a scaling limit of the O'Connell-Yor polymer model.  
When both $N$ and $t$ are large with its ratio $N/t$ fixed, 
it has been known that the polymer free energy $F_N(t)$ defined below~\eqref{ppf}
is proportional to $N$
on average and the fluctuation around the average is of order 
$N^{1/3}$~\cite{MO2007,SV2010}. Furthermore recently it has 
been shown in~\cite{BC2014} that the limiting distribution of the free energy fluctuation under the $N^{1/3}$
scaling is the GUE Tracy-Widom distribution~\cite{TW1994}.
This type of the limit theorem has been obtained also for other models 
related to the O'Connell-Yor model~\cite{BC2015p,BCR2013,CSS2015,FV2013p,OO2014,V2014p}. 
These results reflect the strong universality known as the 
KPZ universality class. 

Although we expect that the same result on the Tracy-Widom asymptotics
can be obtained from our representation~\eqref{cor4},
we consider another scaling limit where the partition function goes to the solution
to the stochastic heat equation (SHE) (or equivalently, the free energy goes to 
the solution to the Kardar-Parisi-Zhang (KPZ) equation).  This scaling limit to
the KPZ equation
has also been known to be universal although in a weaker sense compared with 
the KPZ universality stated above~\cite{AlKhQu2014,BG1997,CT2015p}. 
The height distribution of the KPZ equation has been obtained for a droplet initial data 
in~\cite{ACQ2010, SS2010a,SS2010b,SS2010c,SS2010d}. Since then,
explicit forms of the height distribution have been given for the KPZ equation
and related models for a few initial data~\cite{BC2014,BCF2012,BCFV2014p, CLD2011, IS2012, IS2013, LC2012, OQR2014p, OQR2015p}. In particular for the O'Connell-Yor model~\eqref{ppf}, 
the limiting distribution of the polymer free energy has been obtained by applying the saddle point method to the kernel~\eqref{Lkernel}~\cite{BC2014,BCF2012}.

In this section, we confirm that a similar saddle point analysis can be applicable 
to our biorthogonal kernel~\eqref{kernel}. Since our kernel has  a simple form, we find that
the nontrivial part of this problem reduces only to the asymptotic analyses of the functions
$\psi_k(x;t)$~\eqref{psi} and $\phi_k(x;t)$~\eqref{phi}.

\subsection{The O'Connell-Yor polymer model and the KPZ equation}
Before discussing the saddle point analysis, let us briefly review the scaling limit
to the KPZ equation. 
Hereafter we will write out explicitly the dependence on $\beta$
of the polymer partition function~\eqref{ppf}  as $Z_{N,\beta}(t)$.

Let $\tilde{Z}_{j,\beta}(t):=e^{-t-\beta^2t/2}{Z}_{j,\beta}(t),~j=1,\cdots,N$. 
By It{\^o}'s formula,  we easily find that it satisfies the stochastic differential equations 
\begin{align}
d\tilde{Z}_{j,\beta}(t)=\left(\tilde{Z}_{j-1,\beta}(t)-\tilde{Z}_{j,\beta}(t)\right)dt+\beta \tilde{Z}_{j,\beta}(t)
dB_j(t),
\label{ddf}
\end{align}
where we set $\tilde{Z}_{0,\beta}(t)=0$ and interpret the second term of this equation as It{\^o} type. Now let us take the 
diffusion scaling for~\eqref{ddf}: we set
\begin{align}
t=TM,~~
N=TM-X\sqrt{M}
\label{KPZs1}
\end{align}
and at the same time we scale $\beta$ as
\begin{align}
\beta=M^{-1/4},
\label{1/41}
\end{align}
then take the large $M$ limit. 
the scaling exponent $-1/4$ in~\eqref{1/41}
is known to be universal: it characterize the disorder regime referred to as the intermediate
disorder regime~\cite{AlKhQu2014}, which lies between
weak and strong disorder regimes in directed polymer models in random media  
in $1+1$ dimension.

This $M^{-1/4}$ scaling
can be explained in the following heuristic way.
Let $B_{j}(t),~j=1,\cdots,N$ be $N$ independent one dimensional  
standard Brownian motions. For $N_1,N_2\in \{1,2,\cdots,N\}$, we have
\begin{align}
\langle B_{N_1}(t)B_{N_2}(t)\rangle =t \delta_{N_1,N_2},
\end{align}
where $\langle\cdot\rangle$ represents the expectation value with respect to the
Brownian motions. Now we consider its large $M$ limit under the same scaling
as~\eqref{KPZs1} i.e. $t=MT$ and
\begin{align}
N_k=TM-X_k\sqrt{M},~k=1,2.
\label{tjksc}
\end{align}
Noting that $\lim_{M\rightarrow\infty}\sqrt{M}\delta_{N_1,N_2}=\delta(X_1-X_2)$ under~\eqref{tjksc},
we see 
\begin{align}
\lim_{M\rightarrow\infty}M^{-1/2}\langle B_{N_1}(t)B_{N_2}(t)\rangle=
T\delta(X_1-X_2).
\end{align}
This suggests in a heuristic sense, 
\begin{align}
\lim_{M\rightarrow\infty}M^{-1/4}B_{N_k}(t)= \int_0^Tds\,\eta(s,X_k),~k=1,2. 
\label{1/42}
\end{align}
Here $\eta(T,X)$ with $T>0$ and $X\in\R$
is the space-time white noise with mean $0$ and $\delta$-function covariance,
\begin{align}
\langle\eta(T,X)\rangle=0, ~~\langle\eta(T,X)\eta(T',X')\rangle
=\delta(T-T')\delta(X-X').
\label{wcondition}
\end{align}
Thus considering~\eqref{1/42}, we choose the scaling of $\beta$~\eqref{1/41}.

Under the scaling~\eqref{KPZs1} and~\eqref{1/41}, the following limiting property is
established.
\begin{align}
\lim_{M\rightarrow\infty}\sqrt{M}\tilde{Z}_{N,\beta}(t)
=\Zm(T,X).
\label{CSHE1}
\end{align}
Here $\Zm(T,X)$ is the solution to the SHE 
with the $\delta$-function initial condition,
\begin{align}
&\frac{\partial}{\partial T}\Zm(T,X)=\frac{1}{2}\frac{\partial^2}
{\partial X^2}\Zm(T,X)+\eta(T,X)\Zm(T,X)
\label{SHE}
\\
&\Zm(0,X)=\delta(X),
\label{SHEi}
\end{align}
where $\eta(T,X)$ 
is the space-time white noise with mean $0$ and $\delta$-function 
covariance~\eqref{wcondition}.
The SHE~\eqref{SHE} is known to be well-defined if we interpret 
the multiplicative noise term as It{\^o}-type~\cite{BC1995,M1991}. 
Using this equation, the solution to the KPZ equation can be defined
via 
\begin{align}
h(T,X)=\log (\Zm(T,X)),
\label{cht}
\end{align}
which is called the Cole-Hopf solution to the KPZ equation.
Recently a new regularization for the KPZ equation was developed in~\cite{Ha2013}
(see also \cite{Ku2014p}). 

According to~\cite{BC2014},  a rigorous 
estimate about the convergence 
to the SHE~\eqref{CSHE1}  has been obtained for the O'Connell-Yor model~\cite{MFQu2014p} based on the results 
in~\cite{AlKhQu2014}.  
This type of convergence has been discussed also in interacting 
particle processes~\cite{BG1997,CT2015p}.
For reference we offer a sketch of the derivation 
of~\eqref{CSHE1}. 
For this purpose, we provide the following lemma,

\begin{lemma}~~
For $\tilde{Z}_{N,\beta}(t)$~defined above~\eqref{ddf}, one has
\begin{align}
\tilde{Z}_{N,\beta}(t)
=
\sum_{k=0}^{\infty}\beta^k\sum_{1\le N_1\le\cdots\le N_k\le N}
\int_{\Delta_k(0,t)}
\prod_{j=1}^kdB_{N_j}(t_j)
\cdot
\prod_{j=1}^{k+1}Po(t_j-t_{j-1},N_{j}-N_{j-1})
\label{122}
\end{align}
where $Po(t,n):=e^{-t}t^n/n!$ denotes the Poissonian density  and 
$N_0=1, N_{k+1}=N, s_0=t_0=0, s_N=t_{k+1}=t$. 
$\Delta_n(s,t)$ denotes the region of the integration $s<t_1<\dots<t_n<t$ and 
the It{\^o} integrals on RHS, referred to as the multiple It{\^o} integrals~\cite{It1951,Ma2014},
are performed in time order (i.e. the order of $t_1,\cdots,t_N$).
\end{lemma}

\smallskip
\noindent
{\bf Proof.}
By the definition of $Z_N(t)$~\eqref{ppf}, we have
\begin{align}
\tilde{Z}_{N,\beta}(t)=e^{-t}\int_{0<s_1<\cdots<s_{N-1}<t}
\prod_{j=1}^{N-1}ds_j
\cdot
\prod_{j=1}^Ne^{\beta\left(B_j(s_j)-B_j(s_{j-1})-\frac{\beta(s_j-s_{j-1})}{2}\right)},
\label{Zexpand}
\end{align}
with $s_0$=0 and the integrand of RHS is expressed as
\begin{align}
&\prod_{j=1}^Ne^{\beta\left(B_j(s_j)-B_j(s_{j-1})-\frac{\beta(s_j-s_{j-1})}{2}\right)}
=\prod_{j=1}^N\left(1+e^{\beta\left(B_j(s_j)-B_j(s_{j-1})-\frac{\beta(s_j-s_{j-1})}{2}\right)}-1\right)\notag\\
&=\sum_{m=0}^{\infty}\sum_{1\le M_1<\cdots<M_m\le N}\prod_{j=1}^m
\left(e^{\beta\left(B_{M_j}(s_{M_j})-B_{M_j}(s_{M_j-1})-\frac{\beta(s_{M_j}-s_{M_j-1})}{2}\right)}-1\right).
\end{align}
Here we use the relation on a one-dimensional standard Brownian motion $B(t)$:
one has for $t>s>0$ and $\beta>0$,
\begin{align}
e^{\beta\left(B(t)-B(s)-\frac{\beta(t-s)}{2}\right)}=\sum_{n=0}^\infty \beta^n\int_{\Delta_n(s,t)}
\prod_{j=1}^n
dB(t_j)
\cdot,
\label{121}
\end{align}
where the It{\^o} integrals on RHS, referred to us the multiple It{\^o} integrals, are performed in time order (i.e. the order of $t_1,\cdots,t_N$)~\cite{It1951,Ma2014}.
Using this, we get
\begin{align}
&~\prod_{j=1}^Ne^{\beta\left(B_j(s_j)-B_j(s_{j-1})-\frac{\beta(s_j-s_{j-1})}{2}\right)}
=\sum_{m=0}^{\infty}\sum_{1\le M_1<\cdots<M_m\le N}
\prod_{j=1}^m\sum_{n_j=1}^{\infty}\beta^{n_j}\int_{\Delta_{n_j}(s_{M_j-1},s_{M_j})}\prod_{\ell=1}^{n_j}dB_{M_j}(t_{M_j,\ell})\notag\\
&=\sum_{k=0}^{\infty}\beta^k\sum_{m=0}^{\infty}\sum_{1\le M_1<\cdots<M_m\le N}
\sum_{\substack{n_1,\cdots,n_m=1 \\n_1+\cdots+n_m=k}}^{\infty}
\prod_{j=1}^m\int_{\Delta_{n_j}(s_{M_j-1},s_{M_j})}\prod_{\ell=1}^{n_j}dB_{M_j}(t_{M_j,\ell})
\end{align}
Substituting this into~\eqref{Zexpand}, and performing the integration on $s_1,\cdots,s_{N-1}$, 
we have
\begin{multline}
\tilde{Z}_{N,\beta}(t)
=\sum_{k=0}^{\infty}\beta^k\sum_{m=0}^{\infty}\sum_{1\le M_1<\cdots<M_m\le N}
\sum_{\substack{n_1,\cdots,n_m=1 \\n_1+\cdots+n_m=k}}^{\infty}\int_{\Delta_k(0,t)}
\prod_{j=1}^m\prod_{\ell=1}^{n_j}dB_{M_j}(t_{M_j,\ell})\\
\times 
e^{-t}\prod_{j=1}^{m+1}\frac{(t_{M_j,1}-t_{M_j-1,n_j})^{M_j-M_{j-1}}}{(M_j-M_{j-1})!}
\label{ezt}
\end{multline}
where we set $M_0=1,~M_{m+1}=N$.
Now we introduce the new variables $N_j,~t_j,~j=1,\cdots,k$ by the relation
\begin{align}
N_{n_1+\cdots+n_{j-1}+\ell}=M_j,~t_{n_1+\cdots+n_{j-1}+\ell}=t_{M_j,\ell}~\text{~for~}
\ell=1,\cdots,n_j,~
j=1,\cdots,m.
\label{ntrel}
\end{align}
Then one has $
dB_{M_j}(t_{M_j,\ell})=dB_{N_{n_1+\cdots+n_{j-1}+\ell}}(t_{n_1+\cdots+n_{j-1}+\ell})$
leading to
\begin{align}
\prod_{j=1}^m\prod_{\ell=1}^{n_j} dB_{M_j}(t_{M_j,\ell})=\prod_{j=1}^kdB_{N_j}(t_j).
\label{nmb}
\end{align}
Further from~\eqref{ntrel}, we have
\begin{align}
e^{-t}\prod_{j=1}^{m+1}\frac{(t_{M_j,1}-t_{M_j-1,n_j})^{M_j-M_{j-1}}}{(M_j-M_{j-1})!}
=\prod_{j=1}^{k+1}e^{-(t_j-t_{j-1)}}\frac{(t_j-t_{j-1})^{N_j-N_{j-1}}}{(N_j-N_{j-1})!}
\label{tmn}
\end{align}
where we set $N_0=1,~N_{k+1}=N$.
Substituting these~\eqref{nmb} and~\eqref{tmn} into~\eqref{ezt} and noting
the summations $\sum_{m=0}^{\infty}\sum_{1\le M_1<\cdots<M_m\le N}
\sum_{\substack{n_1,\cdots,n_m=1 \\n_1+\cdots+n_m=k}}^{\infty}$ can be summarized
as the simple form $\sum_{1\le N_1\le\cdots\le N_k\le N}$,
we obtain~\eqref{122}.

\qed

\smallskip

Note that under the scaling~\eqref{KPZs1}, the Poissonian density $Po(t,N)$ goes to the Gaussian density
$g(T,X)=\exp({-{X^2}/{2T}})/{\sqrt{2\pi T}}$, i.e.
\begin{align}
\lim_{M\rightarrow\infty}\sqrt{M}Po(t,N-1)= g(T,X).
\label{pog}
\end{align}
Furthermore by Theorems 4.3 and 4.5 in~\cite{AlKhQu2014}, for a function 
$f(t_1,\cdots,t_k, N_1,\cdots,N_k)$ that converges to 
${\mathfrak{f}}(u_1,\cdots,u_k;y_1,\cdots,y_k)$ under the scaling $t_i=u_i M$ and 
$N_i=u_iM-y_i\sqrt{M},~i=1,\cdots,k$, we have
\begin{align}
&\lim_{M\rightarrow\infty}\frac{1}{M^{3k/4}}\sum_{1\le N_1\le\cdots\le N_k\le N}
\int_{\Delta_k(0,t)}
\prod_{j=1}^k
dB_{N_j}(t_j)
\cdot
f(t_1,\cdots,t_k; N_1,\cdots,N_k)\notag\\
&=\int_{\Delta_k(0;T)}
\prod_{j=1}^k du_j
\cdot
\int_{\R^k}
\prod_{\ell=1}^k
dy_j 
\cdot
\prod_{m=1}^k\eta(t_m,y_m)\cdot \mathfrak{f}(u_1,\cdots,u_k;y_1,\cdots,y_k)
\label{cwhite}
\end{align}
where $\eta(t,y)$ is the space-time white noise with the $\delta$-covariances~\eqref{wcondition}.
Thus from~\eqref{122},~\eqref{pog} and~\eqref{cwhite}, we have under the scaling~\eqref{KPZs1},
\begin{align}
&~\lim_{M\rightarrow\infty}\sqrt{M}\tilde{Z}_{N,\beta}(t)\notag\\
&=\lim_{M\rightarrow\infty}\sum_{k=0}^{\infty}(\beta M^{1/4})^k
\frac{1}{M^{3k/4}}
\sum_{1\le N_1\le\cdots\le N_k\le N}
\int_{\Delta_k(0,t)}
\prod_{j=1}^k
dB_{N_j}(t_j)
\notag\\
&\hspace{8.5cm}\times\prod_{j=1}^{k+1}M^{1/2}Po(t_j-t_{j-1},N_{j}-N_{j-1})\notag\\
&=\sum_{k=0}^\infty
\int_{\Delta_k(T)}
\prod_{j=1}^k
dt_j
\cdot
\int_{\R^k}\prod_{j=1}^k dy_j\cdot \prod_{m=1}^k\eta(t_m,y_m)\cdot\prod_{\ell=1}^{k+1}g(t_{\ell}-t_{\ell-1},y_{\ell}-y_{\ell-1}),
\end{align}
where $t_0=0,t_{k+1}=T,y_0=0,y_{k+1}=X$. Since we easily find that RHS 
of this equation is the solution of the SHE with $\delta$-function initial 
data~\eqref{SHE}, we obtain~\eqref{CSHE1}.

\subsection{The asymptotics of the kernel}
In~\cite{BC2014}, Borodin and Corwin discussed the asymptotics of the Fredholm 
determinant~\eqref{bcrep} under the scaling limit to the KPZ equation, especially
the limiting property of the kernel~\eqref{Lkernel} based on the saddle point
method. Here we check that a similar saddle point method is applicable to our biorthogonal
kernel~\eqref{kernel}. 
The scaling limit we consider is ~\eqref{CSHE1} discussed above, but here we 
adopt its rephrased version stated in~\cite{BC2014}, 
\begin{align}
\lim_{N\rightarrow\infty}\frac{Z_{N,\b=1}(t=\sqrt{TN}+X)}{C(N,T,X)}=\Zm(T,X),
\label{CSHE}
\end{align}
where $C(N,T,X)$ is
\begin{align}
C(N,T,X):=\exp\left(N+\frac{\sqrt{TN}+X}{2}+X\sqrt{\frac{N}{T}}\right)
\left(\frac{T}{N}\right)^{\frac{N}{2}},
\label{cntx}
\end{align}
which is more suitable for our purpose.
To see the equivalence between~\eqref{CSHE1} and~\eqref{CSHE}, we rewrite 
the relation~\eqref{CSHE1} as
\begin{align}
\lim_{N\rightarrow\infty}\beta^{-2}\tilde{Z}_{N,\beta}(t)=\Zm(T,X),
\label{CSHE2}
\end{align}
where we scale $t,~\beta$ as
\begin{align}
t=N+X\sqrt{\frac{N}{T}},~
\beta=\left(\frac{N}{T}\right)^{-1/4}.
\label{KPZs2}
\end{align}
Furthermore  focusing on the scaling property of the partition function
$Z_{N,\beta}(t)=Z_{N,1}(\beta^2 t)
/\beta^{2(N-1)}$, we find 
\begin{align}
\beta^{-2}\tilde{Z}_{N,\beta}(t)=\frac{1}{\beta^{2N}e^{t+\beta^2t/2}}
Z_{N,1}(\beta^2 t)
\end{align}
in distribution. Noticing under the scaling~\eqref{KPZs2}
\begin{align}
\beta^2t=\sqrt{TN}+X,~
\beta^{2N}e^{t+\beta^2t/2}
=C(N,T,X), 
\end{align}
where $C(N,T,X)$ is defined in~\eqref{cntx}, we find that~\eqref{CSHE2} 
is equivalent to~\eqref{CSHE}.

For the moment generating function, ~\eqref{CSHE} implies
\begin{align}
\lim_{N\rightarrow\infty}\E\left(e^{-e^{-u}Z_{N,1}\left(\sqrt{TN}+X\right)}\right)
=\E\left(e^{-e^{-u'}\Zm(T,X)}\right)
=\E\left(e^{-e^{-u'+h(T,X)}}\right),
\label{osk}
\end{align}
where on LHS, $u$ is set to be 
\begin{align}
u=u'+\log C(N,T,X),~
\label{uscaling1}
\end{align}
with $C(T,N,X)$~\eqref{cntx}, and  in the last equality in~\eqref{osk} we used~\eqref{cht}. 
The notions of the
KPZ universality class tell us that the fluctuation of the height
$h(T, X)$ and the position $X$ are scaled as $T^{1/3}$ and $T^{2/3}$ respectively 
for large $T$. Considering them, we set 
\begin{align}
h\left(T,2\gamma_T^2Y\right)=-\frac{\gamma_T^3}{12}+\gamma_T(\tilde{h}(T,Y)-Y^2),
\label{sheight}
\end{align}
where $\gamma_T=(T/2)^{1/3}$. The first term $-\gamma_T^3/12=-T/24$
represents the macroscopic growth with a constant velocity.
The height fluctuation is expressed as
$\tilde{h}(T,Y)$ and the term $Y^2$ reflects the fact that the SHE with
delta-function initial data in~\eqref{SHEi} corresponds to 
the parabolic growth in the KPZ equation~\cite{SS2010b,SS2010c,ACQ2010}. 
Thus substituting $u'=\gamma_ts-\gamma_T^3/12-\gamma_TY^2$, 
$X=2\gamma_T^2Y$ into~\eqref{uscaling1}, we arrive at the 
modified scaling
\begin{align}
u=\gamma_Ts-\frac{\gamma_T^3}{12}-\gamma_TY^2+N+\frac{\sqrt{TN}+2\gamma_T^2Y}
{2}+2\gamma_T^2Y\sqrt{\frac{N}{T}}+\frac{N}{2}\log\frac{T}{N}.
\label{utscaling}
\end{align}
Hence \eqref{osk} is rewritten as
\begin{align}
\lim_{N\rightarrow\infty}\E\left(e^{-e^{-u}Z_{N,1}\left(\sqrt{TN}+2\gamma_t^2Y\right)}\right)
=\E\left(e^{-e^{\gamma_t(\tilde{h}(T,Y)-s))}}\right).
\label{osk2}
\end{align}
with the scaling ~\eqref{utscaling}. This is the scaling limit of the moment generating 
function from the O'Connell-Yor polymer to the KPZ equation. 

It has been known that RHS of this equation
can be represented as 
the Fredholm determinant~\cite{CLDR2010,D2010,D2010p2},
\begin{align}
\E\left(e^{-e^{\gamma_T(\tilde{h}(T,Y)-s)}}\right)=\det
\big(1-\K_{\text{KPZ}}\big)_{L^2(\R)},
\label{KPZrelation}
\end{align}
where the kernel $\K_{\text{KPZ}}(\xi_1,\xi_2)$ is expressed as
\begin{align}
\K_{\text{KPZ}}(\xi_1,\xi_2)=\frac{e^{\gamma_T(\xi_1-s)}}{e^{\gamma_T(\xi_1-s)}+1}
\int_{0}^{\infty}d\lambda\, \Ai(\xi_1+\lambda)\Ai(\xi_2+\lambda).
\label{KPZkernel}
\end{align}
Note that $Y$ does not appear in RHS of this equation.
This kernel first appeared in the studies of the KPZ equation for the narrow wedge initial 
condition \cite{ACQ2010,SS2010b,SS2010a,SS2010c,SS2010d}. 
From the relation~\eqref{KPZrelation} we readily get  
the distribution of the scaled height $\tilde{h}(T,Y)$ given in~\eqref{sheight}. 

By combining the formula~\eqref{cor4} for the O'Connell-Yor polymer 
and the limiting relation~\eqref{osk2} from the O'Connell-Yor polymer to the KPZ equation,
we can obtain~\eqref{KPZrelation} by showing 
\begin{align}
\lim_{N\rightarrow\infty}\det\left(1-\bar{f}_uK\right)_{L^2(\R)}
=\det\left(1-\K_{\text{KPZ}}\right)_{L^2(\R)}
\label{limitFred}
\end{align}
under~\eqref{utscaling}. This was indeed already discussed in \cite{BC2014} by 
using the kernel~\eqref{Lkernel} . 
Here we show that the kernel~\eqref{KPZkernel} appears rather easily from the 
scaling limit of our biorthogonal kernel~\eqref{cor4}.
Using the saddle point method,
we get the following:
\begin{proposition}\label{asyker}
\begin{align}
\lim_{N\rightarrow\infty}\bar{f}_u(x_1)K(x_1,x_2;\sqrt{TN}+2\gamma_T^2Y)=
e^{\frac{\gamma_T}{2}(\xi_1-\xi_2)}
\K_{\rm{KPZ}}(\xi_1,\xi_2).
\label{spresult}
\end{align}
Here the kernel is expressed in terms of $\phi_k(x_1;t)$ and $\psi_k(x_2;t)$ defined by~\eqref{phi} 
and~\eqref{psi} respectively as
\begin{align}
\bar{f}_u(x_1)K(x_1,x_2;t)=\frac{e^{x_1-u}}{e^{x_1-u}+1}\sum_{k=0}^{N-1}\phi_k(x_1;t)\psi_k(x_2;t),
\label{fKkernel}
\end{align}
and we set $u$ to be~\eqref{utscaling} and 
\begin{align}
x_i=\gamma_T\xi_i-\frac{\gamma_T^3}{12}-\gamma_TY^2+N+
\frac{(TN)^{1/2}+2\gamma_T^2Y}{2}+2\gamma_T^2Y\sqrt{\frac{N}{T}}
+\frac{N}{2}\log \frac{T}{N}.
\label{xscaling}
\end{align}
\end{proposition}
\smallskip
\noindent
Since
the factor $e^{\frac{\gamma_T}{2}(\xi_1-\xi_2)}$ in~\eqref{spresult} does not contribute to the Fredholm determinant, we get ~\eqref{limitFred} (though for a complete proof one has to 
prove the convergence of the Fredholm determinant itself, not only the kernel). 
Note that~\eqref{fKkernel} has a similar
structure to the kernel~\eqref{khermite} in the GUE random matrices.
When we discuss certain large $N$ limits in the GUE such as the bulk and the edge
scaling limit, the nontrivial step reduces to the scaling limit of the Hermite polynomial
in~\eqref{khermite}. The same thing happens in our case: the only nontrivial 
step for getting~\eqref{spresult} is the asymptotics of the functions $\psi_k(x;t)$~\eqref{psi} 
and  $\phi_k(x;t)$~\eqref{phi}.
Based on the saddle point method, we obtain the following results of which the proof 
is given in Appendix~\ref{A}.
\begin{lemma}
\begin{align}
\lim_{N\rightarrow\infty}
\frac{\gamma_T}{C(N)}\psi_k(x_i;t)=\lim_{N\rightarrow\infty}\frac{N^{1/2}C(N)}{(2\gamma_T)^{1/2}}\phi_k(x_i;t)={\Ai(\xi_i-\lambda)},
~~i=1,2,
\label{psiphiairy}
\end{align}
where we set $x_i$ as \eqref{xscaling} and $k$ and $t$ as  
\begin{align}
k=N+\frac{N^{1/2}}{(2\gamma_T)^{1/2}}\lambda,~t=\sqrt{TN}+2\gamma_T^2Y.
\label{xkscaling}
\end{align}
The constant $C(N)$ is represented as
$
C(N)=e^{\sum_{j=1}^5C_j}
$ in terms of $C_1,\cdots,C_5$ defined by~\eqref{C1},~\eqref{C24} 
and~\eqref{C5} in Appendix~\ref{A}.
\end{lemma}

\smallskip
On the other hand, when we  take 
the same limit for the other  representation~\eqref{bcrep},  
we can apply the saddle point analysis also to the 
kernel~\eqref{Lkernel} and can get the limiting kernel.
But since it does not correspond to the kernel~\eqref{KPZkernel} directly,
we need an additional step to show the equivalence between the Fredholm determinant
with the limiting kernel and that with~\eqref{KPZkernel}
 (see Sec. 5.4.3 in~\cite{BC2014}).

\smallskip
\noindent
{\bf Proof of Proposition~\ref{asyker}.}
Combining the estimate~\eqref{psiphiairy}
with the simple fact
\begin{align}
\frac{e^{x_i-u}}{e^{x_i-u}+1}=\frac{e^{\gamma_T(\xi_i-s)}}{e^{\gamma_T(\xi_i-s)}+1},~i=1,2,
\end{align}
under~\eqref{utscaling} and~\eqref{xscaling}, we immediately obtain the result~\eqref{spresult}. 
\qed

\section{Conclusion}
For the O'Connell-Yor directed random polymer model, we have established the 
representation (\ref{main})
of the moment generating function for the partition function 
in terms of a determinantal function which is regarded as a one-parameter 
deformation of the eigenvalue density function of the GUE random matrices. 

There are some special mathematical structures behind 
the O'Connell-Yor model which play a crucial role in deriving 
the relation. The first one has been the determinantal representation~\eqref{lemma3} which
is essentially the one with the Sklyanin measure in~\cite{O2012}. 
Next we have introduced another determinantal measure
in enlarged degrees of freedom~\eqref{taweight}. 
Our main theorem has been readily obtained from a simple fact 
about two marginals of this measure (Theorem~\ref{mar}).

We can regard our approach as a generalization of
the one in~\cite{Wa2007} which retains its determinantal structures.
To see this we needed to reinterpret the dynamics in the Gelfand-Tsetlin cone 
introduced in~\cite{Wa2007}
using the weight~\eqref{RR} supported on the partially ordered space $V_N$~\eqref{vn}.
Our approach is a natural generalization of~\cite{Wa2007} from this viewpoint.
It would be an interesting future problem to find a clear relation 
with the Macdonald process~\cite{BC2014}, which is another
generalization of~\cite{Wa2007}.

Applying familiar techniques in random matrix theory to the main result, we have readily 
obtained the Fredholm determinant representation of the moment generating function
whose kernel is expressed as the biorthogonal functions both of which 
are simple deformations of the Hermite polynomials. The asymptotics of the kernel 
under the scaling limit to the KPZ equation can 
be estimated easily by the saddle point analysis.

\appendix
\section{Proof of Lemma~\ref{lem9}.}\label{B}
\begin{figure}[h]
\begin{center}
\includegraphics[scale=0.6]{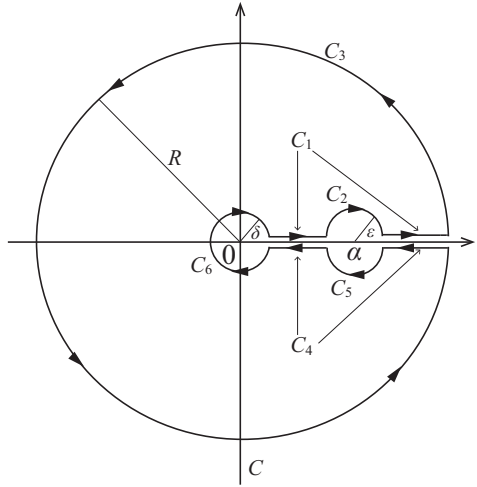}
\caption{\label{contour}
The contour $C$ on $\C$, where $\a\in (0,\infty)$. 
It consists of the paths $C_1,\cdots,C_6$.
}
\end{center}
\end{figure}

First we give a proof of~\eqref{r1}. For this purpose, it is sufficient to 
show the case of $m=1,~x=0$,
\begin{align}
\int_{-\infty}^{\infty}dx\, e^{-ax}f_B(x)=\frac{\pi}{\beta}\cot\left(\frac{\pi a}{\beta}\right).
\label{rr1}
\end{align}
Furthermore setting $e^{\beta x}=y$, $a/\beta=b$, one sees that~\eqref{rr1} is rewritten as
\begin{align}
\int_0^\infty
dy\,
 h_b(y)=\pi\cot\pi b, 
\label{r1'}
\end{align}
where $h_b(y)=y^{-b-1}/(y-1)$ and we take the branch cut of $h_b(y)$
to be the positive real axis. Hence here we prove~\eqref{r1'}. 
We set the contour $C$ as depicted in Fig.~\ref{contour} with $\a=1$.
From the 
Cauchy integral theorem, we find
\begin{align}
\int_{C}
dy\,
h_b(y)=0.
\label{cint0}
\end{align}
Dividing the contour $C$ into $C_i,~i=1,\cdots,6$ as in Fig.~\ref{contour},
we find that
by simple calculations,
\begin{align}
&\lim_{\scriptstyle \delta\rightarrow 0, \e\rightarrow 0 \atop R\rightarrow\infty} 
\int_{C_1}
dy\,
h_b(y) =-\lim_{\scriptstyle \delta\rightarrow 0, \e\rightarrow 0 \atop R\rightarrow\infty} e^{2\pi i b}\int_{C_4}dy\, h_b(y)
=\int_0^\infty dy\, h_b(y),\notag\\
&\lim_{\e\rightarrow0}\int_{C_2dy\,}dy\, h_b(y)=\lim_{\e\rightarrow0}e^{2\pi i b}
\int_{C_5}dy\, h_b(y)=-\pi i,\notag\\
&\lim_{R\rightarrow\infty}\int_{C_3}dy\, h_b(y)=\lim_{\delta\rightarrow 0}
\int_{C_6}dy\, h_b(y)=0,
\label{c36}
\end{align}
where note that the factors $e^{2\pi i b}$s come from the cut locus of $y^{-b}$.
 
From~\eqref{cint0},\eqref{c36}, we get
\begin{align}
0=\lim_{\scriptstyle \delta\rightarrow 0, \e\rightarrow 0 \atop R\rightarrow\infty} \sum_{j=1}^6\int_{C_j}dy\, h_b(y)=(1-e^{-2\pi i b})\int_0^\infty dy\, h_b(y)
\, -(1+e^{-2\pi i b})\pi i,
\end{align}
which leads to~\eqref{r1'}.

Next we give a proof of~\eqref{r2}.  For this purpose we first show the following relation.
Let $I_j(x),~j=1,2,\cdots$, $x\in (0,\infty)$,
be
\begin{align}
I_j(x)=\int_0^\infty dw\, \frac{1}{x-w}\frac{(\log w)^{j-1}}{w+1}.
\label{Ijdef}
\end{align}
Then we have
\begin{align}
I_j(x)=\frac{1}{x+1}\,r_{j}(\log x),
\label{iformula}
\end{align}
where $r_k(x)~(k=0,1,2,\cdots)$  is a $k$th order polynomial of $x$ where
the coefficient of the highest degree is $1/k$.
This relation~\eqref{iformula} can be derived by considering 
the integration of $m_j(w;x):=(\log w)^j/((x-w)(w+1)),~x>0$ with respect to $w$
along the contour $C$ in~Fig.~\ref{contour} with $\a=x$ and $R>1$.

Note that
\begin{align}
\int_{C}dw\, m_j(w;x)=2\pi i\,\frac{(\pi i)^{j}}{x+1},
\label{hcint0}
\end{align}
where RHS corresponds to the residue of $m_j(w;x)$ at $w=-1$.
As in the previous case~\eqref{c36}, one easily gets
\begin{align}
&\lim_{\scriptstyle \delta\rightarrow 0, \e\rightarrow 0 \atop R\rightarrow\infty} \int_{C_1}dw\,m_j(w;x) 
=I_{j+1}(x),\notag\\
&\lim_{\scriptstyle \delta\rightarrow 0, \e\rightarrow 0 \atop R\rightarrow\infty} \int_{C_4}dw\,m_j(w;x)
=-\int_0^\infty dw\,  m_j(we^{2\pi i};x)
=-\int_0^\infty dw\,  \frac{(\log w+2\pi i)^j}{(x-w)(w+1)}\notag\\
&\hspace{4.3cm}=-\sum_{k=0}^{j}\binom{j}{k}(2\pi i)^{j-k}I_{k+1}(x),\notag\\
&\lim_{\e\rightarrow0}\int_{C_2}
dw\,
m_j(w;x)=\pi i\,\frac{(\log x)^j}{x+1},~~
\lim_{\e\rightarrow0}\int_{C_5}m_j(w;x)dw=\pi i\,\frac{(\log x+2\pi i)^j}{x+1},\notag\\
&\lim_{R\rightarrow\infty}\int_{C_3}
dw\,
m_j(w;x)
=\lim_{\delta\rightarrow 0}
\int_{C_6}dw\,m_j(w;x)=0.
\label{hc36}
\end{align}
Substituting~\eqref{hc36} into~\eqref{hcint0}, we find
\begin{align}
2\pi i\frac{(\pi i)^j}{x+1}&=\lim_{\scriptstyle \delta\rightarrow 0, \e\rightarrow 0 \atop R\rightarrow\infty}\sum_{k=1}^6\int_{C_k}dw\, m_j(w;x)\notag\\
&=I_{j+1}(x)-\sum_{k=0}^j\binom{j}{k}(2\pi i)^{j-k}I_{k+1}(x)+\frac{\pi i}{x+1}
\left((\log x)^j+\left(\log x+2\pi i\right)^j\right).
\end{align}
Thus we obtain
\begin{align}
I_{j}(x)=\frac{(\log x)^{j}+(\log x+2\pi i)^j-2(\pi i)^j}{2j(x+1)}-\frac{1}{j}\sum_{k=0}^{j-2}\binom{j}{k}(2\pi i)^{j-1-k}I_{k+1}(x)
\label{irec}
\end{align}
which leads to~\eqref{iformula}.

Here we show~\eqref{r2}. We find that~\eqref{r2} is rewritten as 
\begin{align}
J_B^{*(m)}J_F(x)=q_m\left(\frac{\log x}{\beta}\right)J_F(x),
\label{hformula}
\end{align} 
where $q_m(x)$ is defined below~\eqref{r2} and the functions  $J_F(x)$ and $J_B(x)$ on $\R_+$ are defined by
$J_F(x)=1/(x+1)$ and $J_B(x)=1/\beta x$.

We prove (\ref{hformula}) by using (\ref{iformula}) and by mathematical induction: 
suppose that~\eqref{hformula} holds for 
$m=N-1$. Then 
we get
\begin{align}
J^{*(N)}_BJ_F(x)&=\frac{1}{\beta}\int_{0}^\infty dy\,\frac{1}{x-y}J^{*(N-1)}_BJ_F(y)
=\frac{1}{\beta^N(N-1)!}I_{N}(y)+\sum_{k=0}^{N-2}\frac{c_k}{\beta^{k+1}}I_{k+1}(y)
\label{jbjf}
\end{align}
where $c_k (k=0,1,\cdots,N-2)$ is the coefficient of 
$x^k$ in $q_{N-1}(x)$
and in the last equality
we used the assumption for the mathematical induction and~\eqref{Ijdef}. 
Considering~\eqref{iformula}, we arrive at~\eqref{hformula}.
\qed

\section{Proof of Lemma~\ref{vgt2}.}\label{pvgt}
To show Lemma~\ref{vgt2}, we will use
the following identity. For $(x_1,\cdots, x_{N-1})\in\R^{N-1}$ satisfying $x_1>\cdots>x_{N-1}$ and 
$(y_1,\cdots,y_N)\in\R^N$, we have
\begin{align}
\sum_{\sigma\in S_N}\sgn\sigma \prod_{j=2}^N1_{>0}\left(x_{j-1}-y_{\sigma(j)}\right)
=\sum_{\sigma\in S_N}\sgn\sigma \prod_{j=2}^N1_{>0}\left(x_{j-1}-y_{\sigma(j)}\right)
1_{>0}\left(y_{\sigma(j-1)}-x_{j-1}\right).
\label{i2}
\end{align}
where $S_N$ is the permutation of  $(1,2,\cdots,N)$.
For the proof of~\eqref{i2}, it is sufficient to show for $m=1,2, \cdots, N$,
\begin{align}
&\sum_{\sigma\in S_N}\sgn\sigma\prod_{j=2}^N1_{>0}(x_{j-1}-y_{\sigma(j)})\cdot
\prod_{k=2}^{m}1_{>0}(y_{\sigma(k-1)}-x_{k-1})\cdot 1_{>0}(x_{m}-y_{\sigma(m)})
=0,
\label{i1}
\end{align}
where, as in~\eqref{i2}, we assume the condition $x_1>x_2>\cdots>x_{N-1}$.
This can easily be obtained by noting that
\begin{align}
&\sum_{\sigma\in S_N}\sgn\sigma\prod_{j=2}^N1_{>0}(x_{j-1}-y_{\sigma(j)})\cdot
\prod_{k=2}^{m}1_{>0}(y_{\sigma(k-1)}-x_{k-1})\cdot 1_{>0}(x_{m}-y_{\sigma(m)})\notag\\
&=\sum_{\sigma\in S_N}\sgn\sigma\prod_{\substack{j=2\\j\neq m}}^N1_{>0}(x_{j-1}-y_{\sigma(j)})\cdot
\prod_{k=2}^{m}1_{>0}(y_{\sigma(k-1)}-x_{k-1})\cdot 1_{>0}(x_{m}-y_{\sigma(m)})\notag\\
&=\sum_{\sigma\in S_N}\sgn\sigma\hspace{-5mm}\prod_{\substack{j=2\\j\neq m,~m+1}}^N
\hspace{-5mm}
1_{>0}(x_{j-1}-y_{\sigma(j)})\cdot
\prod_{k=2}^{m}1_{>0}(y_{\sigma(k-1)}-x_{k-1})\cdot 
1_{>0}(x_{m}-y_{\sigma(m)})1_{>0}(x_{m}-y_{\sigma(m+1)})\notag\\
&=0
\end{align}
where in the first equality we used the fact that the factor $1_{>0}(x_{m-1}-y_{\sigma(m)})$
can be omitted in this equation thanks to the factor $1_{>0}(x_m-y_{\sigma(m)})$ with the condition $x_{m-1}>x_{m}$ and the last equality follows from the fact that
the term with $\sigma$ cancels the term with $\sigma'$ where $\sigma'$ is defined in terms of $\sigma$ as $\sigma'(m)=\sigma(m+1)$ and 
$\sigma'(m+1)=\sigma(m)$ with $\sigma'(k)=\sigma(k)$ for $k\neq m,~m+1$.

Using~\eqref{i1}, we have for $x_1>x_2>\cdots>x_{N-1}$
\begin{align}
&\sum_{\sigma\in S_N}\sgn\sigma \prod_{j=2}^N1_{>0}\left(x_{j-1}-y_{\sigma(j)}\right)
=
\sum_{\sigma\in S_N}\sgn\sigma \prod_{j=2}^N1_{>0}\left(x_{j-1}-y_{\sigma(j)}\right)\cdot
(1-1_{>0}(x_{1}-y_{\sigma(1)}))\notag\\
&=\sum_{\sigma\in S_N}\sgn\sigma \prod_{j=2}^N1_{>0}\left(x_{j-1}-y_{\sigma(j)}\right)\cdot
1_{>0}(y_{\sigma(1)}-x_1)\notag\\
&=\sum_{\sigma\in S_N}\sgn\sigma \prod_{j=2}^N1_{>0}\left(x_{j-1}-y_{\sigma(j)}\right)\cdot
1_{>0}(y_{\sigma(1)}-x_1)(1-1_{>0}(x_{2}-y_{\sigma(2)}))\notag\\
&=\sum_{\sigma\in S_N}\sgn\sigma \prod_{j=2}^N1_{>0}\left(x_{j-1}-y_{\sigma(j)}\right)\cdot
\prod_{k=1}^2
1_{>0}(y_{\sigma(k)}-x_k).
\label{i12}
\end{align}
where for the first and the third equality, we used~\eqref{i1} with $m=1$ and $m=2$ respectively.
Performing the procedure in~\eqref{i12} repeatedly, we arrive at~\eqref{i2}.

Now we give a proof of the lemma by the mathematical induction.
The case $N=1$ in is trivial. Suppose that it holds for $N-1$. Then
noticing 
\begin{align}
1_{V_N}(\underline{x}_N)=\prod_{k=2}^N\prod_{j=2}^k1_{>0}\left(x^{(k-1)}_{j-1}-x^{(k)}_j\right)
=1_{V_{N-1}}(\underline{x}_{N-1})\prod_{j=2}^N1_{>0}\left(x^{(N-1)}_{j-1}-x^{(N)}_j\right),
\end{align}
we see that LHS of~\eqref{evgt}  is written as
\begin{align}
&~\sum_{\sigma^{(j)}\in S_j,j=1,\cdots,N}\sgn\sigma^{(N)}\,
1_{V_{N-1}}(\underline{x}_{N-1}^{\sigma})
\prod_{j=2}^N
1_{>0}\left(x^{(N-1)}_{\sigma^{(N-1)}(j-1)}-x^{(N)}_{\sigma^{(N)}(j)}\right)\notag\\
&=\sum_{\sigma^{(j)}\in S_j,j=1,\cdots,N}\sgn\sigma^{(N-1)}\,
1_{V_{N-1}}(\underline{x}_{N-1}^{\sigma})\cdot\sgn\sigma^{(N)}
\prod_{j=2}^N
1_{>0}\left(x^{(N-1)}_{j-1}-x^{(N)}_{\sigma^{(N)}(j)}\right)\notag\\
&=\sum_{\sigma^{(j)}\in S_j,j=1,\cdots,N}\sgn\sigma^{(N-1)}\,
1_{\text{GT}}(\underline{x}_{N-1}^{\sigma})\cdot\sgn\sigma^{(N)}
\prod_{j=2}^N
1_{>0}\left(x^{(N-1)}_{j-1}-x^{(N)}_{\sigma^{(N)}(j)}\right)\notag\\
&=\sum_{\sigma^{(j)}\in S_j,j=1,\cdots,N}\sgn\sigma^{(N)}\,
1_{\text{GT}}(\underline{x}_{N-1}^{\sigma})
\prod_{j=2}^N
1_{>0}\left(x^{(N-1)}_{\sigma^{(N-1)}(j-1)}-x^{(N)}_{\sigma^{(N)}(j)}\right),
\label{evgt2}
\end{align}
where in the second equality we used the assumption for $N-1$.
Note that in the rightmost side of~\eqref{evgt2}, 
the condition $x^{(N-1)}_{\sigma^{(N-1)}(1)}>x^{(N-1)}_{\sigma^{(N-1)}(2)}>\cdots>
x^{(N-1)}_{\sigma^{(N-1)}(N-1)}$ holds for the support of $1_{\text{GT}}(\underline{x}_{N-1}^{\sigma})$.
Thus we can apply~\eqref{i2} to the rightmost side.
We see that it becomes
\begin{align}
&~\sum_{\sigma^{(j)}\in S_j,j=1,\cdots,N}\sgn\sigma^{(N)}\,
1_{\text{GT}}(\underline{x}_{N-1}^{\sigma})
\prod_{j=2}^N
1_{>0}\left(x^{(N-1)}_{\sigma^{(N-1)}(j-1)}-x^{(N)}_{\sigma^{(N)}(j)}\right)
1_{>0}\left(x^{(N)}_{\sigma^{(N)}(j-1)}-x^{(N-1)}_{\sigma^{(N-1)}(j-1)}\right)\notag\\
&=
\sum_{\sigma^{(j)}\in S_j,j=1,\cdots,N}\sgn\sigma^{(N)}
\,
1_{\rm{GT}}(\underline{x}_N^\sigma),
\end{align}
which completes the proof of Lemma~\ref{vgt2}.

%
%
%
%

\qed

\section{The saddle point analysis of $\psi_k(x;t)$}\label{A}
In this Appendix, we give a proof of~\eqref{psiphiairy} based on the saddle point 
method in a similar way to Sec.~5.4.3 in~\cite{BC2014}.
Here we deal with the case of general $Y$ while the case of $Y=0$ was considered
in~\cite{BC2014}.
We focus mainly on the limit about $\psi_k(x;t)$~\eqref{psi} in~\eqref{psiphiairy}
since the case $\phi_k(x;t)$~\eqref{phi} can also be estimated in a 
parallel way. Changing the variable as $w=-i\sqrt{N}z$,~\eqref{psi} 
becomes
\begin{align}
\psi_k(x,t)=\frac{\sqrt{N}}{2\pi i}\int_{-i\infty}^{i\infty}
dz\, e^{f_N(z;t,x)},
\label{psifN}
\end{align}
where
\begin{align}
f_N(z;t,x)=-\sqrt{N}zx+N\frac{z^2t}{2}+(k-N)\log(\sqrt{N}z)
-N\log\Gamma (\sqrt{N} z).
\label{fN}
\end{align}
Substituting~\eqref{xscaling} and~\eqref{xkscaling} into~\eqref{fN}, 
we arrange the first three terms in ascending order of powers of 
$N$ as
\begin{align}
&~-\sqrt{N}zx_i+N\frac{z^2t}{2}+(k-N)\log(\sqrt{N}z)\notag\\
&=N^{3/2}\log N \cdot\frac{z}{2}
+N^{3/2}\left(-z+\frac{T^{1/2}z^2}{2}-\frac{z}{2}\log T\right)
+N\left(\gamma_T^2Yz^2-2\gamma_T^2YT^{-1/2}z-\frac{T^{1/2}z}{2}\right)\notag\\
&~~+N^{1/2}\log N\cdot\frac{\lambda}{2(2\gamma_T)^{1/2}} 
+N^{1/2}\left(\frac{\gamma_T^3}{12}z-\gamma_T\xi_i z
+\gamma_TY^2z-\gamma_T^2Yz
+\frac{\lambda}{(2\gamma_T)^{1/2}}\log z\right).
\label{fN123}
\end{align}
Using the Stirling formula
\begin{align}
\log\Gamma(n)=n\log n-n-\frac{\log2\pi n}{2}+\frac{1}{12n}+O(n^{-3})
\end{align}
for the last term in~\eqref{fN}, we have
\begin{align}
-N\log\Gamma (\sqrt{N}z)
&=-N^{3/2}\log N\cdot \frac{z}{2}+
N^{3/2}\left(z-z\log z\right)+\frac{N}{4}\log N+
N \frac{\log2\pi z}{2}\notag\\
&~~~~-N^{1/2}\frac{1}{12z}+O(N^{-1/2}).
\label{fN4}
\end{align}
Thus from~\eqref{fN123} and~\eqref{fN4}, $f_N(z)$~\eqref{fN} 
can be expressed as
\begin{align}
&f_N(z;t,x)=N^{3/2}f(z)+Ng(z)+N^{1/2}h(z)+C_1+O(N^{-1/2}),
\label{fNex}
\\
&f(z)=\frac{T^{1/2}z^2}{2}-z\log z-\frac{z}{2}\log T,
\label{fz}
\\
&g(z)=-\frac{T^{1/2}z}{2}+\frac{\log z}{2}+\gamma_T^2Yz^2-\frac{2\gamma_T^2Y}{T^{1/2}}z,
\label{gz}
\\
&h(z)=-\frac{1}{12z}+\left(\frac{\gamma_T^3}{12}-\gamma_T^2Y-\gamma_T(\xi_i-Y^2)
\right)z
+\frac{\lambda}{(2\gamma_T)^{1/2}}\log z.
\label{hz}
\end{align}
Here $C_1$, which does not depend on $z$ is written as
\begin{align}
C_1=N^{1/2}\log N\cdot\frac{\lambda}{2(2\gamma_T)^{1/2}}+\frac{N}{2}\log 2\pi
\sqrt{N}.
\label{C1}
\end{align}
We note that $f(z)$ above has a double saddle point
$z_c=T^{-1/2}$ such that $f'(z_c)=f''(z_c)=0$. We expand 
$f(z),~g(z),~h(z)$ around $z_c$. Noting
$f'''(z_c)=2\gamma_T^3$, $g'(z_c)=0,~g''(z_c)=2\gamma_T^2Y-\gamma_T^3$, $h'(z_c)=\gamma_T^3/4-\gamma_T^2Y+\gamma_T(\lambda-\xi_i+Y^2)$,
we get
\begin{align}
N^{3/2}f(z)&=N^{3/2}f(z_c)+\frac{N^{3/2}}{3!}f'''(z_c)(z-z_3)^3+
O(N^{3/2}(z-z_c)^4)\notag\\
&=C_2+N^{3/2}\frac{\gamma_T^3}{3}(z-z_c)^3+O(N^{3/2}(z-z_c)^4),
\label{fex}
\\
Ng(z)&=Ng(z_c)+Ng'(z_c)(z-z_c)+N\frac{g''(z_c)}{2}(z-z_c)^2
+O(N(z-z_c)^3)\notag\\
&=C_3-N\left(\frac{\gamma_T^3}{2}-\gamma_T^2Y\right)(z-z_c)^2+O(N(z-z_c)^3),
\label{gex}
\\
N^{1/2}h(z)&=N^{1/2}h(z_c)+N^{1/2}h'(z_c)(z-z_c)+C_4+O(N^{1/2}(z-z_c)^2)
\notag\\
&=C_4+N^{1/2}\left(\frac{\gamma_T^3}{4}-\gamma_T^2Y+\gamma_T(\lambda-\xi+Y^2)
\right)(z-z_c)+O(N^{1/2}(z-z_c)^2),
\label{hex}
\end{align}
where $C_2,~C_3$ and $C_4$ are
\begin{align}
&C_2:=N^{3/2}f(z_c)=N^{3/2}T^{-1/2}/2,~~C_3:=Ng(z_c)=-N
\left(\frac{1+\log T}{2}+\frac{Y}{2\gamma_T}\right), \notag\\
&C_4:=N^{1/2}h(z_c)=-N^{1/2}\left(
\frac{T^{1/2}}{24}+(\gamma_T^2Y+\gamma_T(\xi-Y^2))T^{-1/2}
-\frac{\lambda\log T^{-1/2}}{(2\gamma_T)^{1/2}}
\right).
\label{C24}
\end{align}
Thus from~\eqref{fNex}--\eqref{hex} and under the scaling
$\frac{z'}{\sqrt{N}}=(z-z_c)$, we have
\begin{align}
f_N(z;t,x_i)=\frac{\gamma_T^3}{3}
\left(z'+\frac{Y}{\gamma_T}-\frac{1}{2}\right)^3-\gamma_T(\xi-\lambda)
\left(z'+\frac{Y}{\gamma_T}-\frac{1}{2}\right)
+\sum_{j=1}^5C_j+O\left(N^{-1/2}\right),
\end{align}
where $C_1,\cdots,C_4$ are defined in~\eqref{C1} and~\eqref{C24} and $C_5$ is
\begin{align}
C_5=-\frac{1}{3}\left(Y-\frac{\gamma_T}{2}\right)^3+\left(\xi-\lambda\right)
\left(Y-\frac{\gamma_T}{2}\right).
\label{C5}
\end{align}
Further changing the variable
$z'+Y/\gamma_T-1/2=-iv/\gamma_T$, we obtain
\begin{align}
&f_N(z;t,x_i)=\frac{i}{3}v^3+i(\xi_i-\lambda)v
+\sum_{j=1}^5C_j+O\left(N^{-1/2}\right),
\label{fNex2}
\end{align}
Hence from~\eqref{psifN} and \eqref{fNex2}, we get the limiting form of 
$\psi_k(x_i;t)$
\begin{align}
e^{-\sum_{j=1}^5C_j}\gamma_T \psi_k(x_i;t)\sim
\frac{1}{2\pi}\int_{-\infty}^{\infty}dv\,e^{\frac{i}{3}v^3+i(\xi_i-\lambda)v}
=\Ai(\xi_i-\lambda),
\end{align}
which is nothing but~\eqref{psiphiairy}.

As with~\eqref{psifN}, we rewrite $\phi_k(x;t)$~\eqref{phi}
by the change of variable $v=\sqrt{N}z$,
\begin{align}
\phi_k(x;t)=\frac{1}{2\pi i}\oint dz\, \frac{e^{-f_N(z;t,x)}}{z},
\end{align}
where $f_N(z;t,x)$ is given in~\eqref{fN}. Applying the same techniques as above
to this equation, we get the result for $\phi_k(x;t)$. 

\section*{Acknowledgments}
The work of T.~I.  and T.~S. is supported by 
KAKENHI (25800215) and  
KAKENHI (25103004, 15K05203, 14510499)
respectively.

\providecommand{\bysame}{\leavevmode\hbox to3em{\hrulefill}\thinspace}
\providecommand{\MR}{\relax\ifhmode\unskip\space\fi MR }
\providecommand{\MRhref}[2]{%
  \href{http://www.ams.org/mathscinet-getitem?mr=#1}{#2}
}
\providecommand{\href}[2]{#2}

\end{document}